\documentclass[12pt]{article}

\usepackage{amssymb}
\newcommand{\lsim}   {\mathrel{\mathop{\kern 0pt \rlap
  {\raise.2ex\hbox{$<$}}}
  \lower.9ex\hbox{\kern-.190em $\sim$}}}
\newcommand{\gsim}   {\mathrel{\mathop{\kern 0pt \rlap
  {\raise.2ex\hbox{$>$}}}
  \lower.9ex\hbox{\kern-.190em $\sim$}}}

\newcommand{\sig}{\sigma}
\newcommand{\sigl}{\sigma_\lambda}
\newcommand{\sigep}{\sigma_\epsilon}
\newcommand{\omegl}{\omega_\lambda}
\newcommand{\omegep}{\omega_\epsilon}

\newcommand{\mapright}[1]{\smash{\mathop{\hbox to 1cm{\rightarrowfill}}\limits_{#1}}}
\def\gappeq{\mathrel{\rlap {\raise.5ex\hbox{$>$}}
{\lower.5ex\hbox{$\sim$}}}}
\def\lappeq{\mathrel{\rlap{\raise.5ex\hbox{$<$}}
{\lower.5ex\hbox{$\sim$}}}}

\begin{document}
\input epsf

\pagestyle{empty}
\begin{flushright}
\end{flushright}
\vspace*{5mm}
\begin{center}
\Large{\bf Cosmological evolution in vector-tensor theories of gravity} \\
\vspace*{1cm} \large{\bf Jos\'e Beltr\'an Jim\'enez$^1$}  and
\large{\bf Antonio L. Maroto$^2$} \\
\vspace{0.3cm} \normalsize
Departamento de F\'{\i}sica Te\'orica\\
Universidad Complutense de Madrid\\
28040 Madrid, Spain

\vspace*{1cm} {\bf ABSTRACT}
\end{center}

We present a  detailed study of the cosmological evolution  in
 general vector-tensor
theories of gravity without potential terms. We consider the
evolution of the vector field throughout the  expansion history of
the universe and carry out a classification of models according to
the behavior of the vector field in each cosmological epoch. We
also analyze the case in which the universe is dominated by the
vector field, performing a complete analysis of the system phase
map and identifying those attracting solutions which give rise to
accelerated expansion. Moreover, we consider the evolution in a
universe filled with a pressureless fluid in addition to the
vector field and study the existence of attractors in which we can
have a transition from matter-domination to vector-domination with
accelerated expansion so that the vector field may play the role
of dark energy. We find that the existence of solutions with
late-time accelerated expansion is a generic prediction of
vector-tensor theories and that such solutions typically lead to
the presence of future singularities. Finally, limits from local
gravity tests are used to get constraints on the value of the
vector field at small (Solar System) scales.

\vspace*{5mm} \noindent

\vspace*{0.5cm}


\vspace{2cm}

\noindent \rule[.1in]{8cm}{.002in}

\noindent $^1$E-mail: jobeltra@fis.ucm.es\\
\noindent $^2$E-mail: maroto@fis.ucm.es \vfill\eject

\setcounter{page}{1} \pagestyle{plain}

\newpage

\section{Introduction}
General Relativity (GR) is considered nowadays the standard theory
of the gravitational phenomena. In this theory, the gravitational
interaction is mediated by a pure spin two field which allows to
describe the geometry of the background space-time where the
physical processes take place. In spite of its elegance and
experimental success, the possible existence of other metric
theories of gravity compatible with Solar System experiments as
well as the importance of knowing the true theory of the
gravitational interaction to describe properly relativistic
astrophysical objects motivated long time ago the search for
alternative gravitational theories with new degrees of freedom, in
addition to the metric tensor. The first of these attempts was the
so-called Brans-Dicke theory in which, apart from having the
metric tensor, the gravitational interaction is described by a
scalar field which makes the gravitational constant to vary in
space and time. Since the time when this new theory was proposed
in the early 60's, many other possibilities containing new degrees
of freedom (not necessarily scalars) have been proposed.

In cosmology, the interest in alternatives to GR has grown
recently because of the existence of several problems which cannot
be satisfactorily addressed within the framework of GR without
resorting to unknown components, namely, dark matter and dark
energy. The need of new dark matter particles first appeared in
order to account for the rotation curves of galaxies which seemed
to indicate the existence of some kind of non-luminous matter in
the haloes of such galaxies. However, its presence is also crucial
to explain some cosmological observations as, for instance, the
formation of the structures that we observe today (galaxies,
clusters, etc.).

The other dark component, dark energy, would be  responsible for
the accelerated expansion of the universe first inferred from
observations of distant type Ia supernovae \cite{SN1998} and
subsequently confirmed by more precise supernovae measurements
\cite{SN} as well as other cosmological probes, mainly CMB
temperature power spectrum and Baryon Acoustic Oscillations
\cite{CMB}. Although it is possible to explain the accelerated
expansion by introducing a cosmological constant term in
Einstein's equations, this poses a problem from the theoretical
point of view because its value, according to observations,  is
extremely small compared to the natural scale of gravity set by
Newton's constant. Although this fact would not need to be a
problem, the actual situation is that a theory with two scales
differing in many orders of magnitude does not seem very {\it
natural}, and that is why this problem is usually referred to as
the naturalness problem. For this reason, many models trying to
play the role of dynamical dark energy have been developed during
the last decade and can be broadly classified in two classes: on
one hand, models in which dark energy is a new field to add to the
standard composition of the universe \cite{quintessence} and, on
the other hand, models in which the accelerated expansion would be
an effect of a modified theory of gravity \cite{DGP}.
Nevertheless, this distinction is not always very clear because
some modified gravity models require the introduction of new
fields which, indeed, may play the role of dark energy.

One class of alternative theories of gravity with an extra field,
in addition to the metric tensor, are the so-called vector-tensor
theories, in which the gravitational action is modified by adding
a vector field that is non-minimally coupled to gravity. The study
of vector-tensor theories to describe the gravitational
interaction as alternatives to GR started long time ago with the
works by Will, Nordtvedt and Hellings \cite{WNH} in the early 70's
as candidates to produce preferred frame effects. After these
pioneering works, vector-tensor theories were abandoned because
gravitational experiments seemed to rule out such preferred frame
effects. Moreover, fluctuations of the vector field could be
either timelike or spacelike so that those models were generally
thought to present instabilities. A detailed treatment on this
issue was developed in \cite{viable} for the case without
potential term and in \cite{Armendariz} for the case with a
potential term and it was shown that there is still some room in
the parameter space for stable models. A special class of these
theories reemerged due to the increasing interest in models with
Lorentz violation \cite{LV}. The breaking of Lorentz invariance
was achieved by the presence of a vector field whose norm was
forced to be constant by means of a Lagrange multiplier. Moreover,
some of these models are free of instabilities as it was shown in
\cite{Eling}.

More recently, after the discovery of the accelerated expansion of
the universe, vector-tensor theories supplemented with potential
terms have got much attention as possible candidates for dark
energy \cite{VDEpot}. Nevertheless,  vector-tensor theories even
in the absence of potential terms have also turned out to be
compelling candidates for dark energy \cite{cosmicvector} and they
could even solve the aforementioned naturalness problem that most
dark energy models exhibit. A remarkable example is the own
electromagnetic field. Indeed, it  has been recently shown that
electromagnetic quantum fluctuations generated during an
inflationary epoch at the electroweak scale can give rise to an
effective cosmological constant whose value is in agreement with
the observed value \cite{EMDE}. Finally, vector fields have also
been used for other cosmological purposes as candidates to drive
an inflationary epoch \cite{Vinf}, to generate non-singular
cosmologies \cite{Novello}, as dark matter candidates \cite{VDM}
and as candidates to solve some of the observed anomalies in the
CMB power spectrum \cite{CMBanomalies}.

Given the increasing interest in vector-tensor theories of gravity
in order to understand the wide variety of cosmological problems
explained above, it seems useful to provide a detailed analysis of
the cosmological evolution in a general vector-tensor theory so
that one can know whether a particular theory will be able to play
a determined role in the universe history. In this work, we shall
focus on the problem of dark energy and obtain the necessary
conditions to produce late-time acceleration, although the results
obtained will be general enough to be useful in other cosmological
contexts as well.

The paper is organized as follows: in section 2 we describe the
general vector-tensor theory and give the corresponding equations.
Section 3 is devoted to study the evolution of the vector field in
an isotropic Friedman-Robertson-Walker universe. We shall give the
behavior of the vector field in the different phases that the
universe has undergone, namely: inflation, radiation domination
and matter domination. After that, in section 4 we shall study the
case in which the temporal component of the vector field dominates
the universe and, in section 5, we shall obtain the region in the
parameter space in which we might have attracting solutions with
accelerated expansion. The case of a universe filled with matter
plus the vector field will be studied in section 6 where it is
shown that  solutions exist which describe a transition from
matter-dominance to vector-dominance with accelerated expansion.
In section 7 we use current limits on local gravity tests to
obtain constraints on the vector field at small scales.

\section{Vector-tensor theories of gravity}

We shall start by writing the most general action for a
vector-tensor theory without any other restriction apart from
having second order linear equations of motion \cite{Will}:
\begin{eqnarray}
S[g_{\mu\nu},A_\mu]=\int d^4x\sqrt{-g}\left[-\frac{1}{16\pi
G}R+\omega RA_\mu A^\mu+\tilde{\sigma} R_{\mu\nu}A^\mu
A^\nu+\tau\nabla_\mu A_\nu \nabla^\mu A^\nu+\varepsilon F_{\mu
\nu}F^{\mu\nu}\right],\label{Action}
\end{eqnarray}
with $\omega$, $\tilde{\sigma}$, $\tau$, $\varepsilon$
dimensionless parameters and $F_{\mu\nu}=\partial_\mu
A_\nu-\partial_\nu A_\mu$. In the so-called $\AE$ther-Einstein
models, the vector field norm is fixed by introducing a lagrange
multiplier in the action of the form $\lambda_m(A_\mu A^\mu\pm
m^2)$ so that $A_\mu$ is constrained to be either timelike or
spacelike. In other cases, the vector field is supplied with a
mass term or even more complicated potential terms. As we said
above, throughout this work we shall focus on vector-tensor
theories without lagrange multipliers nor potential terms. Notice
also that terms given in (\ref{Action}) are the only possibilities
without introducing new scales in the theory and that gives rise
to linear equations of motion for the vector field. If such an
action is just a low energy limit of some underlying theory, one
would expect to have corrections involving terms of dimension
higher than 4 which would be suppressed by some scale $M$.

For subsequent calculations we shall work with an alternative form
of action (\ref{Action}) obtained via an integration by parts:
\begin{eqnarray}
S[g_{\mu\nu},A_\mu]=\int d^4x\sqrt{-g}\left[-\frac{1}{16\pi
G}R+\omega RA_\mu A^\mu+\sig R_{\mu\nu}A^\mu
A^\nu+\lambda(\nabla_\mu A^\mu)^2+\epsilon F_{\mu
\nu}F^{\mu\nu}\right]\label{Action2}
\end{eqnarray}
where the new parameters relate to the old ones as follows:
\begin{eqnarray}
\sig&=&\tilde{\sigma}-\tau\nonumber\\
\lambda&=&\tau\nonumber\\
\epsilon&=&\frac{2\varepsilon+\tau}{2}.
\end{eqnarray}
We prefer the new form of the action because it allows a more
suggestive interpretation for each term, namely: the
$\epsilon$-term is nothing but the $U(1)$ gauge invariant kinetic
term for the vector field, the $\lambda$-term is analogous to the
gauge fixing term introduced in the electromagnetic quantization
and, finally, both the $\omega$ and $\sigma$ terms are non-minimal
couplings to gravity and play the role of effective mass terms for
the vector field driven by gravity.

The gravitational equations obtained from action (\ref{Action2})
by varying with respect to the metric tensor can be written in the
following way:
\begin{equation}
G_{\mu \nu}=8\pi G\left(\omega T^\omega_{\mu\nu}+\sig
T_{\mu\nu}^{\sig}+\lambda T^\lambda_{\mu\nu}+\epsilon
T^\epsilon_{\mu\nu}+T^{NG}_{\mu\nu}\right),
\end{equation}
where $T^{NG}_{\mu\nu}$ is the energy-momentum tensor
corresponding to other fields rather than $A_\mu$ (generally the
inflaton, matter and radiation) and we have defined:
\begin{eqnarray}
&T^\omega_{\mu\nu}&=2\left[\Box A^2
g_{\mu\nu}+A^2 G_{\mu\nu}+RA_\mu A_\nu-\nabla_\mu\nabla_\nu A^2\right]\nonumber\\
\nonumber\\
&T^{\sig}_{\mu\nu}&=g_{\mu\nu}\left[\nabla_\alpha\nabla_\beta\left(A^\alpha
A^\beta\right)-R_{\alpha\beta}A^\alpha
A^\beta\right]+\Box\left(A_\mu
A_\nu\right)-2\nabla_\alpha\nabla_{(\mu}\left(A_{\nu)}A^\alpha\right)+4A^\alpha
A_{(\mu}R_{\nu)\alpha}\nonumber\\\nonumber\\
&T^\lambda_{\mu\nu}&=g_{\mu\nu}\left[\left(\nabla_\alpha
A^\alpha\right)^2 +2A^\alpha\nabla_\alpha\left(\nabla_\beta
A^\beta\right)\right] -4A_{(\mu}\nabla_{\nu)}\left(\nabla_\alpha
A^\alpha\right)\nonumber\\\nonumber\\
&T^\epsilon_{\mu\nu}&=4F_{\mu\alpha}F_\nu^{\;\;\alpha}
-g_{\mu\nu}F_{\alpha\beta}F^{\alpha\beta}\nonumber\\\nonumber\\
&T^{NG}_{\mu\nu}&=\frac{2}{\sqrt{-g}}\frac{\delta S_{NG}}{\delta
g^{\mu\nu}}
\end{eqnarray}
with $\Box=\nabla_\mu\nabla^\mu$, $A^2=A_\mu A^\mu$ and brackets
in a pair of indices denoting symmetrization with respect to the
corresponding indices.

Apart from the gravitational equations we can obtain a set of
field equations for $A_\mu$ by varying the action with respect to
the vector field to give:
\begin{eqnarray}
2\epsilon \nabla_\nu F^{\mu\nu}-\lambda\nabla^\mu(\nabla_\nu
A^\nu) +\omega RA^\mu+\sig R_\nu^\mu A^\nu=0.
\end{eqnarray}

As we are interested in the cosmological evolution of the vector
field (especially as candidate for dark energy) we shall focus on
the simplest case in which the field is homogeneous, i.e., we
shall study the evolution of the zero Fourier mode of the vector
field. Actually, this will correspond to all Fourier modes whose
physical wavelengths are much larger than the Hubble radius
(super-Hubble modes). In fact, this is the relevant part of the
field for the cosmological expansion evolution, although the
inhomogeneous part could also be very important for the
anisotropies of the CMB or for structure formation, but this is
out of the scope of this work. Besides, we shall choose the
spatial component of the field lying along the $z$-axis in such a
way that we can write $A_\mu=(A_0(t),0,0,A_z(t))$ and, therefore,
we have axial symmetry around that axis. Thus, the metric tensor
will be appropriately described by that of the axisymmetric
Bianchi type I space-time:
\begin{equation}
ds^2=dt^2-a_\perp(t)^2\left(dx^2+dy^2\right)-a_\parallel(t)^2dz^2.
\end{equation}
where $a_\perp$ and $a_\parallel$ are the transverse and
longitudinal scale factors respectively. For this metric, the
field equations read:
\begin{eqnarray}
&&\lambda\left[\ddot{A}_0+(2H_\perp+H_\parallel)\dot{A}_0\right]+
\\&&\left[(2\omega+\sig)(2H_\perp^2+H_\parallel^2)+
(2\omega+\sig+\lambda)(2\dot{H}_\perp+\dot{H}_\parallel)
+2\omega(2H_\perp
H_\parallel+H_\perp^2)\right]A_0=0,\nonumber\\\nonumber\\
&&2\epsilon\left[\ddot{A}_z+(2H_\perp-H_\parallel)\dot{A}_z\right]+\\
&&\left[2\omega(3H_\perp^2+2\dot{H}_\perp)+(2\omega+\sig)(\dot{H}_\parallel+H_\parallel^2
+2H_\perp H_\parallel)\right]A_z=0,\nonumber
\end{eqnarray}
where a dot stands for derivative with respect to the cosmic time
$t$ and $H_\parallel=\dot{a}_\parallel/a_\parallel$ and
$H_\perp=\dot{a}_\perp/a_\perp$ are the longitudinal and
transverse expansion rates respectively. In these equations we see
that the expansion of the universe provides and effective mass for
each component of the vector field as well as a friction term. It
is interesting to note that the dynamics of $A_0$ and $A_z$ are
driven by $\lambda$ and $\epsilon$ respectively, whereas the rest
of parameters of the action, $\omega$ and $\sigma$, only affect
the effective mass of the field. In fact, the presence of
non-vanishing values for $\lambda$ and $\epsilon$ ensures the
existence of evolving $A_0$ and $A_z$ respectively. On the
contrary, if one of these parameters is zero, the corresponding
component does not have dynamics and, in general, will vanish.

The highly isotropic CMB power spectrum that we observe today
shows that the anisotropy at the last scattering surface was very
small so that it is justified to consider small deviations from a
pure isotropic universe so that
\begin{eqnarray}
a_{\perp}(t)&=&a(t)\nonumber\\
a_{\parallel}(t)&=&a(t)\left(1+\frac{1}{2}\;h\right)
\end{eqnarray}
with $h\ll 1$ and $a(t)$ the isotropic scale factor. Then, we can
describe the anisotropy by means of the degree of anisotropy $h$
in terms of which the metric can be written as a perturbed FLRW
given by:
\begin{equation}
ds^2=dt^2-a(t)^2\left(\delta_{ij}+h_{ij}\right)dx^idx^j
\end{equation}
with $h_{ij}=h\delta_{iz}\delta_{jz}$ and $a(t)$ the usual scale
factor. Moreover, the degree of anisotropy can be related to the
linearized Einstein tensor for the Bianchi type I metric as
follows:
\begin{equation}
G_\perp-G_\parallel=\frac{1}{2a^3}\frac{d}{dt}\left(a^3\dot{h}\right)
+\mathcal{O}(h^2)
\end{equation}
where $G_\perp=G^x_x=G^y_y$ and $G_\parallel=G^z_z$. Then, from
Einstein equations we can obtain the evolution for the degree of
anisotropy which happens to be:
\begin{equation}
h=16\pi G\int\frac{1}{a^3}\left[\int
a^3(p_\parallel-p_\perp)dt\right]dt\label{doan}
\end{equation}
In principle, the problem has not been solved yet because the
expression inside the integral will depend on $h$ as well.
However, we can obtain an approximate solution by replacing such
an integrand by its expression in the isotropic case. In fact, one
could obtain more accurate solutions by an iterative process.
Thus, if we now assume that the only source of anisotropy comes
from the spatial component of the vector field we have that:
\begin{eqnarray}
a^2(p_\parallel-p_\perp)&=&\left[(24\omegep+12\omegep\sigep+8\sigep+3\sigep^2)H^2
+(12\omegep+6\omegep\sigep+4\sigep+\sigep^2)\dot{H}\right]A_z^2\nonumber\\
&&+4\sigep HA_z\dot{A}_z-2(2+\sigep)\dot{A}_z^2\label{deltap}
\end{eqnarray}
where $H=\dot{a}/{a}$ the Hubble expansion rate and we have
introduced the notation $\omegep=\omega/\epsilon$ and
$\sigep=\sig/\epsilon$. In the rest of this work we shall perform
a detailed analysis of the isotropic evolution so that we could
eventually use those results to evaluate (\ref{doan}) and,
therefore, to discriminate those models in which the degree of
anisotropy grows or decays as the universe expands, that is, what
models would give rise to large scale anisotropies. For a detailed
treatment on the anisotropy evolution in dark energy models see
\cite{anievol}. Moreover, the generated large-scale anisotropy
would affect the photons coming from the last scattering surface
so that it would give a new contribution to the low multipoles of
the CMB. In fact, this can be used to rule out those models in
which the new contribution is larger than the observed one.

To end this section we would like to comment on a very interesting
feature of these models. In \cite{MDE} it was shown that a dark
energy field carrying a non-vanishing density of momentum could
modify the usual interpretation of the CMB dipole as well as the
value of the quadrupole. The density of momentum of dark energy
can be interpreted as a relative motion of this component with
respect to the others. Given that photons, baryons and dark matter
particles were strongly coupled in the early universe, they
originally shared a common large scale rest frame so that, after
becoming decoupled, they should remain at rest with respect to
each other because of momentum conservation. However, the presence
of a dark energy fluid with a non-vanishing density of momentum
allow the existence of relative velocities among all the
components without violating the momentum conservation. Therefore,
as a vector field has spatial components one would expect it to
carry density of momentum, essentially determined by $A_i$, and,
as consequence, it would be a natural candidate for a moving dark
energy model. Nevertheless, once one uses the equations of motion
it turns out that the density of momentum vanishes identically,
i.e., $T^{0i}=0$ over the field equations. Hence, although the
vector field can provide large scale anisotropy supported by its
spatial component, it does not modify the cosmic rest frame so
that the rest of fluids will generally share a common large scale
rest frame and no effects on the CMB dipole are expected. However,
it is possible that vector perturbations could be supported by
perturbations of the vector field so that, unlike in standard
$\Lambda$CDM, we could obtain large peculiar velocities at large
scales as those observed in \cite{Pecvel}.

\section{Evolution in an isotropic universe}
In this section we shall write down and solve the equations for
the vector field in a universe dominated by an isotropic perfect
fluid, with the energy density of the vector field negligible. In
such a case, both expansion factors become the same
$a_\perp=a_\parallel=a$ as well as the expansion rates
$H_\perp=H_\parallel=H$. This is equivalent to neglect the effects
of the small degree of anisotropy that may be present. In such a
case, the field equations read:
\begin{eqnarray}
\ddot{A}_0+3H\dot{A}_0+\left[3(4\omegl+\sigl)H^2+
3(1+2\omegl+\sigl)\dot{H}\right]A_0&=&0\nonumber\\
\ddot{A}_z+H\dot{A}_z+\frac{1}{2}\left[3(4\omegep+\sigep)H^2
+(6\omegep+\sigep)\dot{H}\right]A_z&=&0.\label{feqRW}
\end{eqnarray}
On the other hand, the energy density associated to $A_0$ and
$A_z$ in the isotropic case are:
\begin{eqnarray}
\rho_{A_0}&=&\lambda\left[3(3+2\omegl+2\sigl)H^2A_0^2
+6(1+2\omegl+\sigl)HA_0\dot{A}_0+\dot{A}_0^2\right]\nonumber\\
\rho_{A_z}&=&-\frac{2\epsilon}{a^2}\left[-(3\omegep+\sigep)H^2A_z^2
+(6\omegep+\sigep)HA_0\dot{A}_0+\dot{A}_0^2\right].\label{rhoI}
\end{eqnarray}
In these expressions we have introduced again the notation
$\omegl=\omega/\lambda$, $\sigl=\sig/\lambda$,
$\omegep=\omega/\epsilon$ and $\sigep=\sig/\epsilon$, which only
makes sense for $\lambda\neq0$ and $\epsilon\neq0$. However, if
that is not the case, the corresponding component does not have
dynamics and, generally, vanishes so that it does not play any
role, as we explained above.

Although we are restricting ourselves to the case of isotropic
expansion, one should be aware of the fact that the presence of a
small shear could modify the evolution of the spatial components
of the vector field because, in the stability analysis around a
FRW metric, there could be vanishing eigenvalues \cite{Barrow}
and, as a consequence, the evolution of $A_z$ could be different
from that determined by (\ref{feqRW}).

In next subsections we shall study the evolution of the vector
field in the different epochs of the expansion history of the
universe and carry out a classification of the models according to
their behaviors.


\subsection{Inflationary (de Sitter) epoch}
During the inflationary era, the universe undergoes an exponential
expansion so that the Hubble parameter $H$ is constant, i.e.,
$a\propto e^{Ht}$. In such a case, the solutions to (\ref{feqRW})
can be expressed as:
\begin{eqnarray}
A_0&=&\left(A_0^+e^{c_0Ht}+A_0^-e^{-c_0Ht}\right)e^{-3Ht/2}\nonumber\\
A_z&=&\left(A_z^+e^{c_zHt}+A_0^-e^{-c_zHt}\right)e^{-Ht/2}\label{Infsol}
\end{eqnarray}
where
\begin{eqnarray}
c_0&=&\frac{1}{2}\sqrt{9-48\omegl-12\sigl}\\
c_z&=&\frac{1}{2}\sqrt{1-24\omegep-6\sigep}
\end{eqnarray}
Then, the evolution of the temporal component depends on whether
$c_0$ is real or complex. That way, we obtain that it oscillates
with frequency $|c_0|$ and it is modulated by a damping factor of
the form $e^{-3Ht/2}$ for $16\omegl+4\sigl>3$ whereas it evolves
as $e^{(c_0-3/2)Ht}$ for $16\omegl+4\sigl< 3$. When
$16\omegl+4\sigl= 3$ we have that $c_0=0$ and the vector field
evolve as $A_0=(C_1^0+C_2^0\;t)e^{-3Ht/2}$.

Concerning $A_z$, it has an oscillating evolution with frequency
$|c_z|$ suppressed by $e^{-Ht/2}$ for $24\omegl+6\sigl>1$ and it
evolves as $e^{(c_z-1/2)Ht}$ for $24\omegl+6\sigl< 1$. Finally,
for $24\omegl+6\sigl=1$ we have that $c_z=0$ and the field evolve
as $A_z=(C_1^z+C_2^z\;t)e^{-3Ht/2}$.

When we insert solutions (\ref{Infsol}) into (\ref{rhoI}) we
obtain the energy density evolutions, which can be written as:
\begin{eqnarray}
\rho_{A_0}&=&\rho_{A_0^+}a^{2c_{0}-3}+\rho_{A_0^-}a^{-2c_{0}-3}\\
\rho_{A_z}&=&\rho_{A_z^+}a^{2c_{z}-3}+\rho_{A_z^-}a^{-2c_{z}-3}
\end{eqnarray}
where we have defined:
\begin{eqnarray}
\rho_{A_0^\pm} &=&\lambda \left[3(3+2\omegl+2\sigl)
+6(1+2\omegl+\sigl)(\pm c_0-3/2)+(\pm c_0-3/2)^2\right](HA_0^\pm)^2\\
\rho_{A_z^\pm} &=&-2\epsilon \left[-(3\omegep+\sigep)
+(6\omegep+\sigep)(\pm c_z-1/2)+(\pm
c_z-1/2)^2\right](HA_z^\pm)^2.
\end{eqnarray}
In these expressions we see that the temporal component is
suppressed during inflation in models with $2c_0<3\Rightarrow
4\omegl+\sigl>0$ whereas an inflationary epoch amplifies $A_0$ for
those models with $4\omegl+\sigl<0$. For $4\omegl+\sigl=0$, the
temporal component has constant energy density. Moreover, for the
aforementioned special case with $c_0=0$ the energy density is
given by:
\begin{equation}
\rho_{A_0}\vert_{c_0=0}=\frac{\lambda}{2}C_2^0
\left[3(5-8\omegl)HC_2^0\;
t-(24\omegl-15)HC_1^0+2C_2^0\right]a^{-3}
\end{equation}
so it decays as $\sim ta^{-3}$ (unless $5-8\omegl=0$).

Analogously, we obtain that the spatial component is amplified
during inflation in models with $12\omegep+3\sigep<-4$, it is
suppressed for those models with $12\omegep+3\sigep>-4$ and it has
constant energy density if the condition $12\omegep+3\sigep=-4$ is
satisfied. Again, for the special case with $c_z=0$ we have a
different evolution given by:
\begin{equation}
\rho_{A_z}\vert_{c_z=0}=\frac{\epsilon}{3}C_2^z
\left[(5-12\omegep)HC_2^z\;
t-(12\omegep-5)HC_1^z+2C_2^z\right]a^{-3}
\end{equation}
so it decays as $\sim ta^{-3}$ (unless $5-12\omegep=0$).

Finally, notice that the oscillating behavior of the field will
translate into an oscillating evolution of the energy density so
that in those cases in which the field oscillates the energy
density for the corresponding component is suppressed by a factor
$a^{-3}$. In fact, if the inflationary era lasts $N$ e-folds,
i.e., the scale factor increases as $a_{end}/a_{in}=e^{N}$, we can
calculate the amplification or suppression of the field at the end
of inflation and is given by:
\begin{eqnarray}
\ln\left[\frac{A_0(t_{end})}{A_0(t_{in})}\right]=
\left[{\textrm{Re}}(c_0)-\frac{3}{2}\right]N\\
\ln\left[\frac{A_z(t_{end})}{A_z(t_{in})}\right]=
\left[{\textrm{Re}}(c_z)-\frac{1}{2}\right]N
\end{eqnarray}
For the energy densities of each component we can proceed
similarly to obtain:
\begin{eqnarray}
\ln\left[\frac{\rho_{A_0}(t_{end})}{\rho_{A_0}(t_{in})}\right]&=&
\left[2{\textrm{Re}}(c_0)-3)\right]N\\
\ln\left[\frac{\rho_{A_z}(t_{end})}{\rho_{A_z}(t_{in})}\right]&=&
\left[2{\textrm{Re}}(c_z)-3)\right]N.
\end{eqnarray}
In these expressions, the real parts of $c_0$ and $c_z$ have to be
taken because they are either real or purely imaginary and, in the
latter case, the only effect of $c_{0,z}$ is to produce
oscillations, but neither suppression nor amplification of the
vector field, as we commented above. In the special cases with
$c_{0,z}=0$ we have to add a term $\ln\frac{t_{end}}{t_{in}}$ on
the right hand sides of the above expressions, although such a
term is usually much smaller than $N$ and can be safely neglected.



\subsection{Barotropic fluid domination}
In a universe dominated by a barotropic perfect fluid with
constant equation of state $w=p/\rho$, the scale factor evolves
according to a power law of the form $a\propto t^p$ with
$p=\frac{2}{3(1+w)}$ so that $H=p/t$. In such a case, the field
equations (\ref{feqRW}) have the following solutions:
\begin{eqnarray}
A_0(t)=A_0^+t^{\alpha_+}+A_0^-t^{\alpha_-}\nonumber\\
A_z(t)=A_z^+t^{\beta_+}+A_z^-t^{\beta_-}\label{solRM}
\end{eqnarray}
with
\begin{eqnarray}
\alpha_\pm &=&\frac{1}{2}\left[1-3p\pm
\sqrt{1+6(1+4\omegl+2\sigl)p+3(3-16\omegl-4\sigl)p^2}\right]\\
\beta_\pm &=&\frac{1}{2}\left[1-p\pm
\sqrt{1+2(6\omegep+\sigep-1)p-(24\omegep+6\sigep-1)p^2}\right]\label{solRMindex}
\end{eqnarray}
As in the inflationary epoch, we see that the components of the
vector field have two different types of evolution depending on
whether $\alpha_\pm$ and $\beta_\pm$ are real or complex. Thus, if
the term inside the root is positive, the corresponding component
of the field will evolve as a power law essentially given by the
growing mode whereas if the term inside the root is negative, the
field will oscillate with an amplitude proportional to
$t^{(1-3p)/2}$ for $A_0$ or $t^{(1-p)/2}$ for $A_z$. In fact, for
$\alpha,\beta\in\mathbb{C}$, the solutions of the vector field can
be expressed as:
\begin{eqnarray}
A_0(t)&=&t^{\footnotesize{\textrm{Re}}(\alpha)}\left[C_1^0\cos\left({\textrm{Im}}(\alpha)\ln
t\right)
+C_2^0\cos\left({\textrm{Im}}(\alpha)\ln t\right)\right]\nonumber\\
A_z(t)&=&t^{\footnotesize{\textrm{Re}}(\beta)}\left[C_1^z\cos\left({\textrm{Im}}(\beta)\ln
t\right) +C_2^z \cos\left({\textrm{Im}}(\beta)\ln t\right)\right]
\end{eqnarray}
where we see that the vector field actually oscillates
harmonically in $\ln t$ and not in the proper time $t$. There is
still another special case which is that when we have degeneration
in the solutions, i.e., when $\alpha_+=\alpha_-$ or
$\beta_+=\beta_-$. If any of these relations takes place, the
corresponding component of the vector field has a logarithmic
solution in addition to the potential solution, being the complete
solution as follows:
\begin{eqnarray}
A_0&=&\left(C_1^0+C_2^0\ln t\right)t^{(1-3p)/2}\nonumber\\
A_z&=&\left(C_1^z+C_2^z\ln t\right)t^{(1-p)/2}
\end{eqnarray}

The evolution of the energy densities are achieved by inserting
solutions (\ref{solRM}) into (\ref{rhoI}) and, for
$\alpha,\beta\in\mathbb{R}$, are given by:
\begin{eqnarray}
\rho_{A_0}=\rho_{A_0^+}a^{\kappa_{0}^+}+\rho_{A_0^-}a^{\kappa_0^-}\\
\rho_{A_z}=\rho_{A_z^+}a^{\kappa_z^+}+\rho_{A_z^-}a^{\kappa_z^-}
\end{eqnarray}
where:
\begin{eqnarray}
\rho_{A_0^\pm} &=&\lambda \left[3(3+2\omegl+2\sigl)p^2
+6(1+2\omegl+\sigl)p\;\alpha_\pm+\alpha_\pm^2\right](A_0^\pm)^2\\
\rho_{A_z^\pm} &=&-2\epsilon \left[-(3\omegep+\sigep)p^2
+(6\omegep+\sigep)p\;\beta_\pm+\beta_\pm^2\right](A_z^\pm)^2
\end{eqnarray}
and
\begin{eqnarray}
\kappa_{0}^\pm&=&2\frac{\alpha_\pm-1}{p}\nonumber\\
\kappa_{z}^\pm&=&2\frac{\beta_\pm-1-p}{p}\label{kappabar}
\end{eqnarray}
Again, given that $\kappa_{0,z}^+\geq\kappa_{0,z}^-$, the energy
densities will evolve proportionally to $a^{\kappa_{0,z}^+}$ so
that the vector field behaves as the superposition of two perfect
fluids with equations of state
$w_{0,z}=-\frac{1}{3}(1+\kappa_{0,z}^+)$. However, in some cases,
it might happen that $\rho_{A_{0,z}^+}$ vanishes and, as a
consequence, the corresponding energy density will evolve as
$a^{\kappa_{0,z}^-}$ rather than $a^{\kappa_{0,z}^+}$.

When $\alpha_\pm$ and $\beta_\pm$ are complex, the energy
densities still evolve as
$\rho_{A_{0,z}}=\rho_{A_{0,z}^+}a^{\kappa^+_{0,z}}$ although we
must replace $\alpha_+\rightarrow {\textrm{Re}}(\alpha_+)$ and
$\beta_+\rightarrow {\textrm{Re}}(\beta_+)$ in (\ref{kappabar})
and $\rho_{A_{0,z}^+}$ are oscillating functions instead of
constants.

For the degenerate case with a logarithmic solution, the energy
densities are given by:
\begin{eqnarray}
\rho_{A_0}\vert_{\alpha_+=\alpha_-}=\lambda\frac{(C_2^0)^2}{2(1-p)}}
\left[\left[1-2p+(24\omegl-15)p^2\right]\left(\frac{C_1^0}{C_2^0}+\ln
t\right)+2(1-p)\right]a^{\kappa_0\\
\rho_{A_z}\vert_{\beta_+=\beta_-}=\epsilon\frac{(C_2^z)^2}{(1-3p)}
\left[\left[-1+6p+(12\omegep-5)p^2\right]\left(\frac{C_1^z}{C_2^z}+\ln
t\right)+2(3p-1)\right]a^{\kappa_z}
\end{eqnarray}
Therefore, the evolution is the same as in the non-degenerate case
modified by a logarithmic variation.

In any case, we can carry out a classification of the models
according to whether the energy density of each component grows,
decays or remains constant. Finally, we can also identify scaling
behaviors or whether the energy density of the vector field grows
or decays with respect to that of the dominant component. In Fig.
\ref{A0clasification} and Fig. \ref{Azclasification} we show the
evolution in the different regions in the parameter space.

\subsubsection{Radiation dominated epoch}

In the radiation dominated epoch we have that $p=1/2$ and the
evolution of the vector field, according to (\ref{solRMindex}), is
given by:
\begin{eqnarray}
\alpha_\pm^R&=&-\frac{1}{4}\left[1\mp\sqrt{25+12\sigl}\right]\\
\beta_\pm^R&=&\frac{1}{4}\left[1\pm\sqrt{1-2\sigep}\right]
\end{eqnarray}
These expressions allow to obtain that the temporal component
evolves as a power law for $\sigl>-\frac{25}{12}$ whereas it
oscillates with an amplitude decaying as $t^{-1/4}$ for $\sigl
<-\frac{25}{12}$. The degenerate case happens for
$\sigl=-\frac{25}{12}$. For the spatial component of the vector
field, we have power law behavior for $\sigep<\frac{1}{2}$ and it
oscillates with an amplitude growing as $t^{1/4}$ for
$\sigep>\frac{1}{2}$, whereas the logarithmic solution appears for
$\sigep=\frac{1}{2}$. Notice that the evolution of the vector
field does not depend on the parameter $\omega$ of the action,
which is due to the fact that the Ricci scalar $R$ vanishes in a
universe dominated by radiation.

Concerning the evolution of the energy densities, according to
(\ref{kappabar}), we have that:
\begin{eqnarray}
\kappa^R_{0\pm}&=&-5\pm\sqrt{25+12\sigl}\\
\kappa^R_{z\pm}&=&-5\pm\sqrt{1-2\sigep}
\end{eqnarray}

If we want the vector field to be dominated by its temporal
component we must impose the condition $\kappa_{0+}>\kappa_{z+}$,
which leads to the constraint $6\sigl+\sigep>-12$. That way we
prevent the generation of large scale anisotropy that would be in
conflict with observations.

\begin{figure}[hb!]
\begin{minipage}{1 \textwidth}
\begin{center}
\small{
\begin{tabular}{|c||c|c|c|}
\hline & & &\\
 Property & Inflation  & Radiation & Matter\\
& & &\\
\hline \hline & & &\\
 Oscillating &  $4\omegl+\sigl>\frac{3}{4}$ & $\sigl <-\frac{25}{12}$  &
  $2\omegl-\sigl>\frac{27}{8}$ \\
& & &\\
\hline & & &\\
 Decaying &  $0<4\omegl+\sigl<\frac{3}{4}$  & $-\frac{25}{12}<\sigl<-2$  &
 $3<2\omegl-\sigl<\frac{27}{8}$   \\
& & &\\
\hline & & &\\
 Scaling&  $4\omegl+\sigl=0$  & $\sigl=-2$ & $2\omegl-\sigl=3$  \\
& & &\\
\hline & & &\\
 Growing & $4\omegl+\sigl<0$  &  $\sigl>-2$ & $2\omegl-\sigl<3$  \\
& & &\\
\hline & & &\\
 Constant &  $4\omegl+\sigl=0$  &  $\sigl=0$ & $2\omegl-\sigl=0$  \\
& & &\\
\hline
\end{tabular}}
\end{center}
\end{minipage}
\\
\vspace{0.5cm}
\begin{minipage}{1 \textwidth}
\begin{center}
{\epsfxsize=17cm \epsfbox{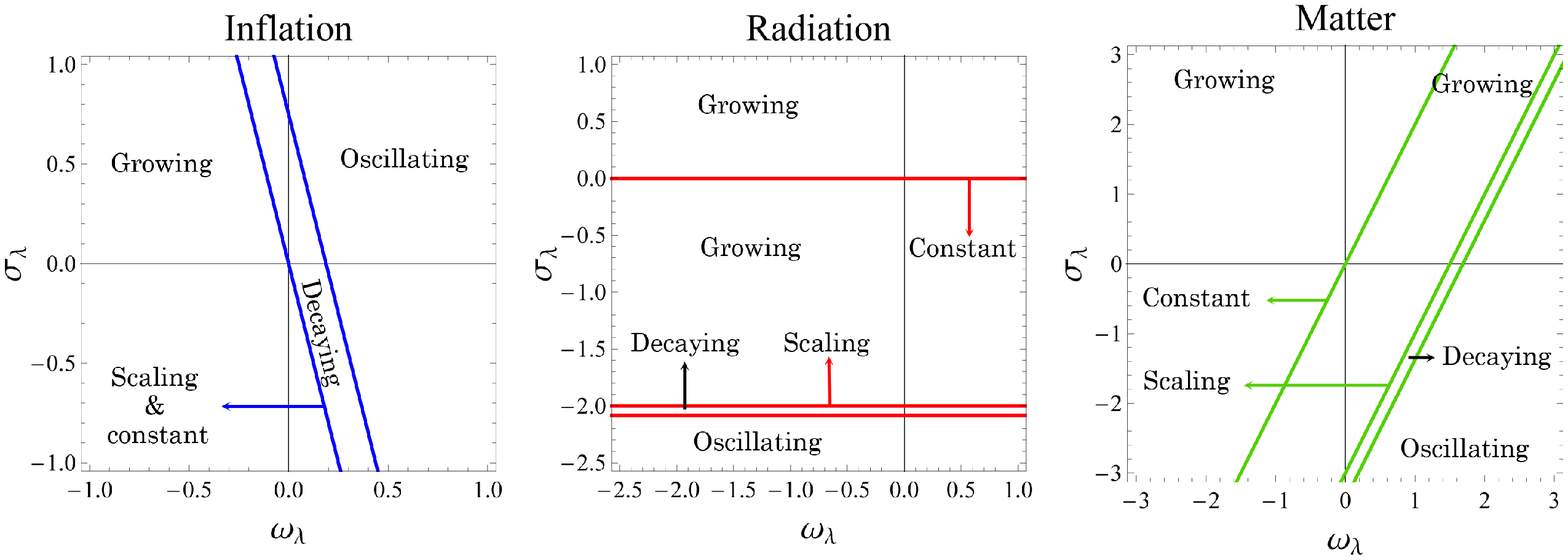}}
\end{center}
\end{minipage}
\caption{\small{Behavior of the temporal component of the vector
field in the different phases of the universe history according to
the values of the parameters of the model. The labels Decaying and
Growing refers to the behavior with respect to the dominant
component, whereas Scaling means that it evolves in the same way
as the background.}}\label{A0clasification}
\end{figure}

The temporal component of the vector field will have constant
energy density in this epoch for $\kappa_{0+}^R=0$ which is
satisfied by the model with $\sigl=0$ and it has scaling evolution
if $\kappa_{0+}^R=-4$ which happens for $\sigl=-2$. From this last
condition we also obtain that, in models with $\sigl>-2$, the
energy density associated to the temporal component grows with
respect to that of the radiation fluid $\rho_R$ whereas in those
models with $\sigl<-2$, the ratio $\rho_{A_0}/\rho_R$ decays as
the universe expands. See Fig. \ref{A0clasification} for a summary
of these behaviors.

Concerning the spatial component, the condition of constant energy
density is satisfied for models with $\sigep=-12$ whereas it
scales as radiation in models with $\sigep=0$. Finally, the energy
density associated to the spatial component of the vector field
grows (decays) with respect to $\rho_R$ in models with $\sigep<0$
($\sigep>0$). This classification is shown in Fig.
\ref{Azclasification}.


\subsubsection{Matter dominated epoch}
In a universe dominated by a pressureless fluid the scale factor
evolves as $a\propto t^{2/3}$ so that
\begin{eqnarray}
\alpha_\pm^M&=&-\frac{1}{6}\left[3\mp\sqrt{81-48\omegl+24\sigl}\right]\\
\beta_\pm^M&=&\frac{1}{6}\left[1\pm\sqrt{1-24\omegep-12\sigep}\right]
\end{eqnarray}
In this case, the temporal component oscillates with amplitude
proportional to $t^{-1/2}$ for $16\omegl-8\sigl>27$, it evolves
with a power law for $16\omegl-8\sigl< 27$ and the degenerate case
corresponds to $16\omegl-8\sigl= 27$. On the other hand, $A_z$
follows a power law evolution for $24\omegep+12\sigep<1$, it
oscillates with amplitude proportional to $t^{1/6}$ for
$24\omegep+12\sigep>1$ and the degenerate case happens for
$24\omegep+12\sigep=1$.

Moreover, $\kappa^R_{0\pm}$ and $\kappa^R_{z\pm}$ become:
\begin{eqnarray}
\kappa^M_{0\pm}&=&\frac{1}{2}\left[-9\pm\sqrt{81-48\omegl+24\sigl}\right]\\
\kappa^M_{z\pm}&=&\frac{1}{2}\left[-9\pm\sqrt{1-24\omegep-12\sigep}\right]
\end{eqnarray}
In this epoch, the condition for the temporal contribution to
dominate over the spatial one reads
$6(\omegep-2\omegl)+3(\sigep-2\sigl)>-80$.

The scaling behavior for the energy density associated to the
temporal component in this epoch is obtained from the condition
$\kappa_{0+}^M=-3$, whose solution is $2\omegl-\sigl=3$. Moreover,
for $2\omegl-\sigl<3$ the energy density of the vector field grows
with respect to that of matter and, as a consequence, the universe
is eventually dominated by it. Notice that this is a necessary
condition to have a dark energy model, i.e., those models with
$2\omegl-\sigl>3$ will never dominate the energy content of the
universe if there is a matter component and, as a consequence, it
cannot be the responsible for the present acceleration. Fig.
\ref{A0clasification} shows this classification.

\begin{figure}[h!]
\begin{minipage}{1 \textwidth}
\begin{center}
\small{
\begin{tabular}{|c||c|c|c|}
\hline & & &\\
 Property & Inflation  & Radiation & Matter\\
& & &\\
\hline \hline & & &\\
 Oscillating &  $4\omegep+\sigep>\frac{1}{6}$ & $\sigep >\frac{1}{2}$  &
  $2\omegep+\sigep>\frac{1}{12}$ \\
& & &\\
\hline & & &\\
 Decaying &  $-\frac{4}{3}<4\omegep+\sigep<\frac{1}{6}$  & $0<\sigep<1$ &
 $-\frac{2}{3}<2\omegep+\sigep<\frac{1}{12}$   \\
& & &\\
\hline & & &\\
 Scaling&  $4\omegep+\sigep=-\frac{4}{3}$  & $\sigep=0$ & $2\omegep+\sigep=-\frac{2}{3}$  \\
& & &\\
\hline & & &\\
 Growing & $4\omegep+\sigep<-\frac{4}{3}$  &  $\sigep<0$ & $2\omegep+\sigep<-\frac{2}{3}$  \\
& & &\\
\hline & & &\\
 Constant &  $4\omegep+\sigep=-\frac{4}{3}$  &  $\sigep=-12$ & $2\omegep+\sigep=-\frac{20}{3}$  \\
& & &\\
\hline
\end{tabular}}
\end{center}
\end{minipage}
\\
\vspace{0.5cm}
\begin{minipage}{1 \textwidth}
\begin{center}
{\epsfxsize=17cm \epsfbox{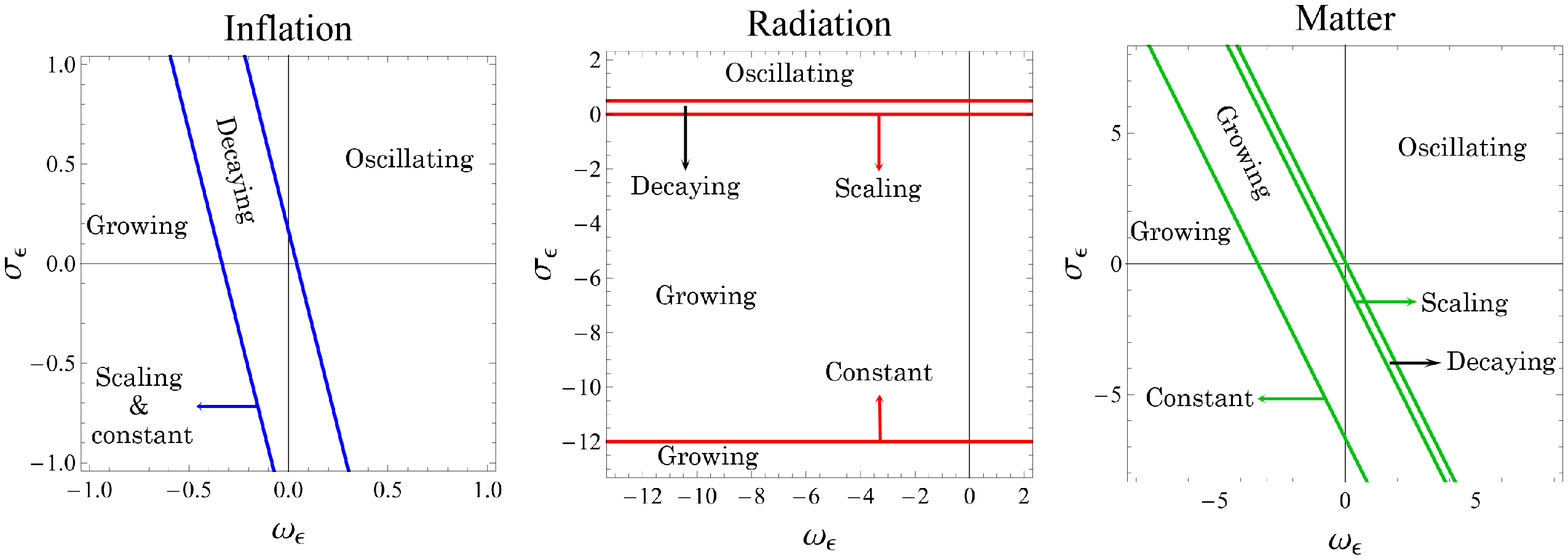}}
\end{center}
\end{minipage}
\caption{\small{Behavior of the spatial component of the vector
field in the different phases of the universe history according to
the values of the parameters of the model. The labels Decaying and
Growing refers to the behavior with respect to the dominant
component, whereas Scaling means that it evolves in the same way
as the background.}}\label{Azclasification}
\end{figure}

For the spatial component we obtain constant energy density in
models with $2\omegep+\sigep=-20/3$ and scaling evolution for
$2\omegep+\sigep=-2/3$. Finally, the energy density of $A_z$ grows
(decays) with respect to $\rho_M$ in models with
$2\omegep+\sigep<-2/3$ ($2\omegep+\sigep>-2/3$). This is shown in
Fig. \ref{Azclasification}.


\section{Vector dominance}

In this section we shall study the case in which the universe
becomes dominated by the temporal component of the vector field so
that the anisotropy is small and we can use the isotropic
equations. Thus, we have the field equation for $A_0$ given in
(\ref{feqRW}) and the two Einstein equations:
\begin{eqnarray}
3H^2&=&8\pi G\rho_{A_0}\label{asympFeq}\\
3H^2+2\dot{H}&=&-8\pi Gp_{A_0}.\label{asympEeq}
\end{eqnarray}
Although, of course, only two of the three equations are
independent, it will be useful to work with all of them.

For the subsequent analysis, it will be convenient to introduce
the field variable $x\equiv \frac{d\ln A_0}{d\ln a}$ so that we
can obtain the following autonomous system:
\begin{eqnarray}
\frac{d
H}{dN}&=&3\frac{(2\omegl+\sigl)x^2-2(4\omegl+\sigl)x-3(4\omegl+\sigl)(2\omegl+\sigl+1)}
{\left[x+3(2\omegl+\sigl+1)\right]^2}H\\
\frac{dx}{dN}&=&-\frac{x^2+6(2\omegl+\sigl+1)x+3(2\omegl+2\sigl+3)}
{x+3(2\omegl+\sigl+1)}x\label{autsys}
\end{eqnarray}
with $N=\ln a$. These two equations can be combined to give the
following equation for the trajectories in the phase map:
\begin{equation}
\frac{dH}{dx}=-3\frac{(2\omegl+\sigl)x^2-2(4\omegl+\sigl)x-3(4\omegl+\sigl)(2\omegl+\sigl+1)}
{\left[x^2+6(2\omegl+\sigl+1)x+3(2\omegl+2\sigl+3)\right]\left[x+3(2\omegl+\sigl+1)\right]}\frac{H}{x}
\end{equation}
This equation can be readily integrated for given values of the
parameters, although we shall not do it, but we shall study the
phase map and, from its features, we shall obtain the relevant
information.

In addition to the equations given above, we have the following
constraint provided by the Friedman equation:
\begin{equation}
\frac{1}{3}\lambda
A_0^2\left[x^2+6(2\omegl+\sigl+1)x+3(2\omegl+2\sigl+3)\right]=1
\end{equation}
This relation constrains the possible values of $x$ because the
condition
\begin{equation}
\lambda\left[x^2+6(1+\sigl+2\omegl)x+3(3+2\sigl+2\omegl)\right]\geq0\label{Adregion}
\end{equation}
must hold. This condition restricts the physically admissible
trajectories as those for which the vector field carries positive
energy density and can always be achieved by means of a suitable
choice of the sign of $\lambda$. Notice that all the dependency of
the problem on the parameter $\lambda$ is indeed contained in this
condition.

\begin{figure}[hb!]
\begin{center}
{\epsfxsize=13cm \epsfbox{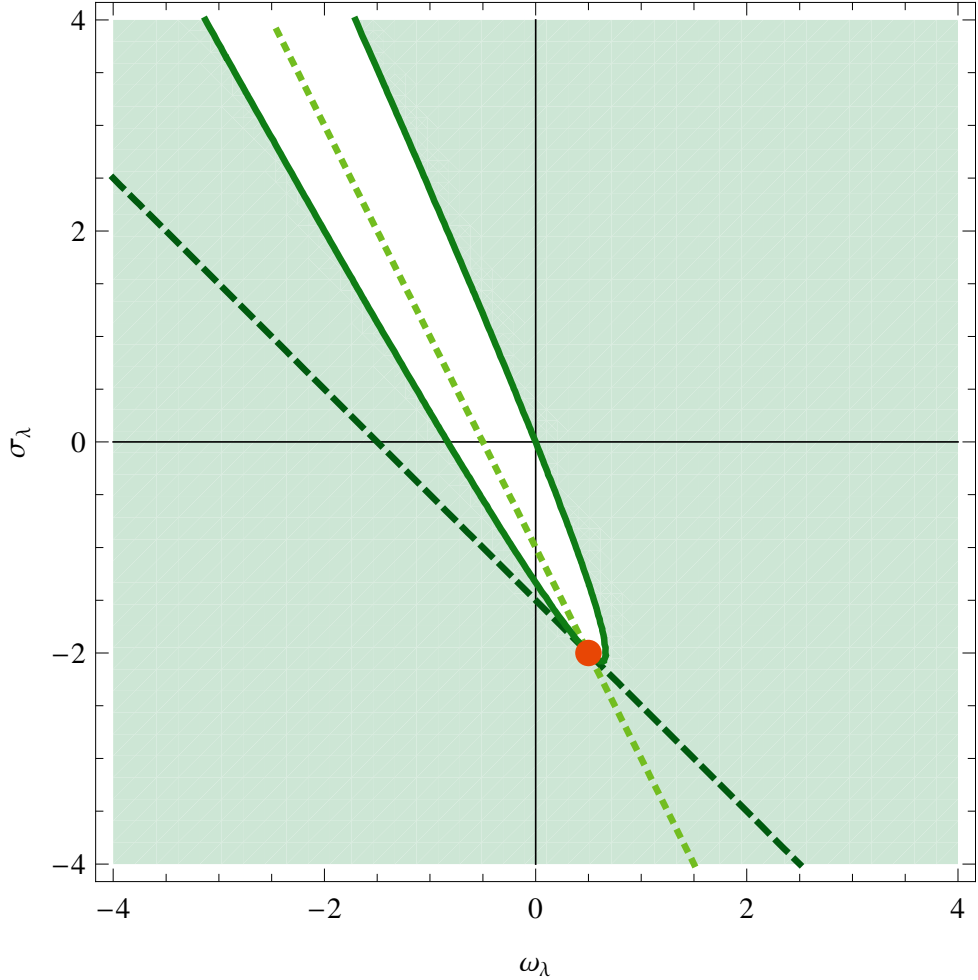}}\caption{\small This plot
shows the different cases explained in the main text in the
parameter space. The green shaded region corresponds to models in
which both $x_\pm$ exist (Case I) whereas in the white region
neither of them is present (Case II). The solid line identifies
the models for which $x_+=x_-$ corresponding to Case III. The
dashed line (whose equation is $2\omegl+2\sigl+3=0$) represents
those models that have either $x_-$ or $x_+$ at the origin (Case
Ib) and the models whose parameters lie on the dotted line (with
equation $2\omegl+\sigl+1=0$) have the separatrix $x_s$ at the
origin. Thus, in the region above (below) that line $x_s$ is
negative (positive). Case Ic corresponds to the piece of the
dotted line inside the white region whereas the piece of the
dotted line inside the grey region corresponds to Case IIb.
Finally, the orange dot with parameters $\omegl=1/2, \sigl=-2$
gives the Case IIIb in which $x_\pm=x_s=0$. }\label{VDregion1}
\end{center}
\end{figure}

\begin{figure}[hb!]
\begin{center}
{\epsfxsize=15cm \epsfbox{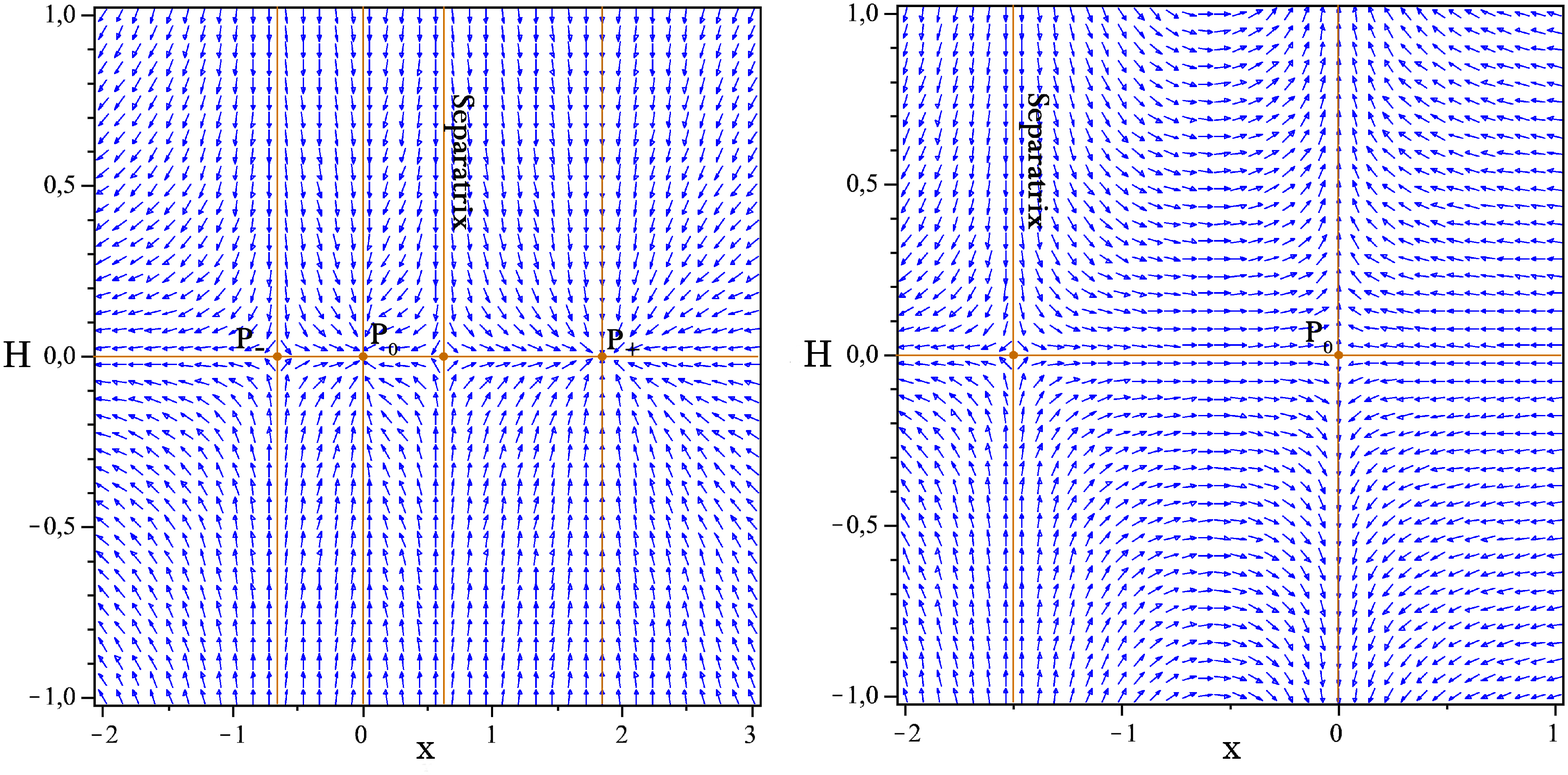}}\caption{\small{These
plots shows two examples of phase maps for the autonomous system
describing a universe dominated by the vector field. The left
panel corresponds to a model lying in the grey region in Fig.
\ref{VDregion1} in which the two critical points $P_\pm$ are
present whereas the right panel shows the phase map for a model in
which these two critical points do not exist. We can see that all
the critical points as well as the separatrix are vertical
tangents and the $x$-axis is a horizontal separatrix, as explained
in the main text.}}\label{phasemap}
\end{center}
\end{figure}

In (\ref{autsys}) we see that the equation $\frac{dH}{dN}=0$ has
two solutions: $H=0$ and $x=x_l^\pm$, with:
\begin{equation}
x^\pm_l=\frac{4\omegl+\sigl\pm
\sqrt{(4\omegl+\sigl)\left[2(5\omegl+2\sigl)+3(2\omegl+\sigl)^2\right]}}{2\omegl+\sigl}.
\end{equation}
whereas the solutions of $\frac{dx}{dN}=0$ are $x=0$ and
$x=x_c^\pm$, with:
 \vspace{0.01cm}
\begin{equation}
x^\pm_c=-3(2\omegl+\sigl+1)\pm\sqrt{6(5\omegl+2\sigl)+9(2\omegl+\sigl)^2}.
\end{equation}
Therefore, the autonomous system has generally three critical
points: $P_0=(0,0)$ and $P_\pm=(x_c^\pm,0)$. Moreover, when the
equality $x_c^\pm=x_l^\pm$ takes place, we have a critical line
instead of a critical point because both $\frac{dH}{dN}$ and
$\frac{dx}{dN}$ vanish regardless the value of $H$. Notice that
the critical points $P_\pm$ only exist when the constraint
$6(5\omegl+2\sigl)+9(2\omegl+\sigl)^2\geq0$ is satisfied, which
corresponds to the white region shown in Fig. \ref{VDregion1}.
Moreover, the trajectories in the phase map have vertical tangents
in the lines $x=0$ and $x=x_c^\pm$ so that, apart from being
critical points, they are vertical separatrices (see Fig.
\ref{phasemap}). However, these are not the only separatrices in
the phase map, but we have another vertical one in
$x_s=-3(2\omegl+\sigl+1)$ as well as one horizontal one in the
horizontal axis $H=0$. One interesting feature of the phase map is
that $x_s=(x_c^++x_c^-)/2$, i.e., the separatrix at $x_s$ is
always located in the middle of the two critical points. Notice
that the critical points $x_c^\pm$ also separate the region with
physically admissible trajectories according to (\ref{Adregion})
which imposes the energy density of the vector field to be
positive, although to identify each of these regions we need to
specify the sign of $\lambda$. According to the previous
discussion, the phase map will be divided in several rectangular
regions parallel to the axis which, indeed, are disconnected from
each other, i.e., the trajectories will not be able to cross from
one to another. However, the particular picture will depend on the
particular values of the parameters, being possible to distinguish
the following cases (see Fig. \ref{phasemap}):

\begin{itemize}
\item Case I: The critical points $x_c^\pm$ exist and are
different from each other, which imposes the condition
$9(2\omegl+\sigl+1)^2-3(2\omegl+2\sigl+3)>0$ and corresponds to
the grey region in Fig. \ref{VDregion1}. Moreover, in this case we
still have three different possibilities:
\begin{itemize}
\item Case Ia: Both critical points are different from zero and
they do not coincide with $x_s$. In this case we have 4 vertical
separatrices and the phase map is divided into 10 disconnected
regions.

\item Case Ib: One of the two critical points is zero. This case
happens when $2\omegl+2\sigl+3=0$ but $2\omegl+\sigl+1\neq0$ so
that they are not zero simultaneously. In fact, if
$2\omegl+\sigl+1$ is positive (negative), then $x_c^+$ ($x_c^-$)
is at the origin. Notice that this simply says that, given that
the separatrix located at $x_s$ is in the middle of $x_-$ and
$x_+$, the critical point that is at the origin depends on the
sign of $x_s$. In this case we only have 3 vertical asymptotes
and, as a consequence, the phase map contains 8 regions. This case
corresponds to the dashed line in Fig. \ref{VDregion1}, which
represents the equation $2\omegl+2\sigl+3=0$

\item Case Ic: The vertical separatrix satisfies $x_s=0$ so that
the two critical points are symmetric with respect to the origin.
The condition $x_s=0$ reads in terms of the parameters
$2\omegl+\sigl+1=0$ and, as a consequence, the critical point
$P_0$ is absent. In this case we also have 3 vertical asymptotes
and 8 regions in the phase map and corresponds to the dotted line
inside the grey region in Fig. \ref{VDregion1}.
\end{itemize}

\item Case II: Neither of the two critical points $x_c^\pm$
exists. This case satisfies
$9(2\omegl+\sigl+1)^2-3(2\omegl+2\sigl+3)>0$ and corresponds to
the white region in Fig. \ref{VDregion1}. Again, we have several
subcases:
\begin{itemize}
\item Case IIa: The separatrix $x_s$ is different from zero. Then,
we have two different vertical asymptotes and the phase map
contains 6 regions.

\item Case IIb: The separatrix is located at the origin, $x_s=0$
so that $2\omegl+\sigl+1=0$ ($P_0$ is not a critical point) and we
are left with just one vertical asymptote that divides the phase
map in 4 regions. This case corresponds to the dotted line inside
the white region in Fig. \ref{VDregion1}.
\end{itemize}

\item Case III: The two critical points become equal so we only
have one critical point which, indeed, coincides with $x_s$, i.e.,
$x_c^+=x_c^-=x_s=-3(2\omegl+\sigl+1)$. This case is represented by
the solid line in Fig. \ref{VDregion1}. The two possibilities we
have in this case are:
\begin{itemize}
\item Case IIIa: The critical point is different from zero so that
we have two vertical asymptotes and the phase map contains 6
regions.

\item Case IIIb: The critical point is zero. Then, we only have
one vertical separatrix and four regions. This case corresponds to
the particular model $(\sigl=-2,\omegl=1/2)$ and is represented by
the orange dot in Fig. \ref{VDregion1}.
\end{itemize}
\end{itemize}

\begin{figure}[ht!]
\begin{center}
{\epsfxsize=17cm \epsfbox{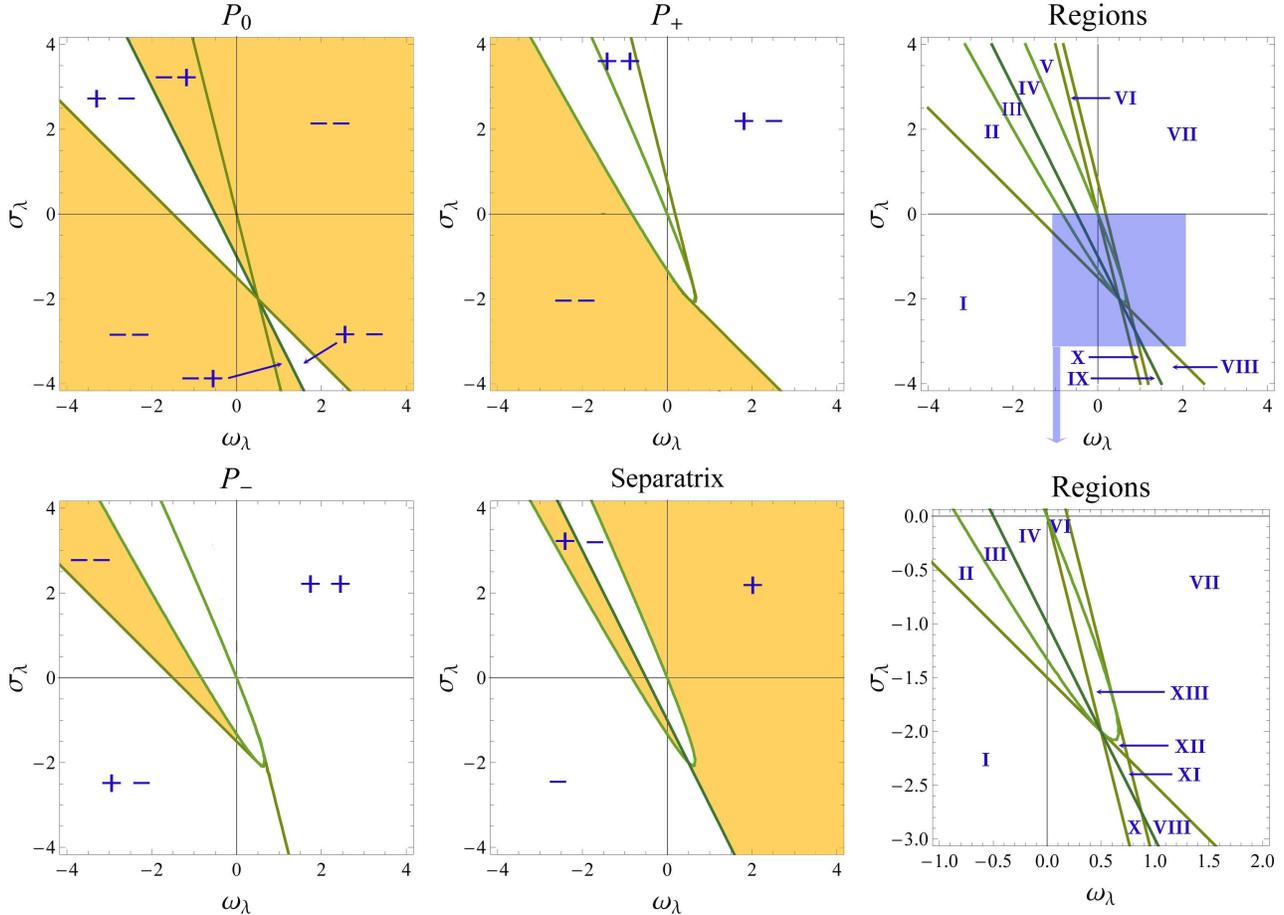}}\caption{\small In
this plot we show the different regions obtained in the parameter
space according to the features of the critical points. We have
shaded (in orange) the regions where the eigenvalue of $x$ is
negative so that the trajectories in the phase map approach the
corresponding point. We have also indicated the sign of each
eigenvalue in the form $(\mu_x,\mu_H)$. For the separatrix we
indicate the regions where the trajectories approach the
separatrix. Finally, the two plots on the right show the 13
regions explained in the text (and summarized in table 1). The
last plot corresponds to a zoom of the blue-shaded region in the
third panel.}\label{criticalpoints1}
\end{center}
\end{figure}

Now that we know the arrangement of the phase map for the
different models attending to the existence and location of the
critical points, we shall study the particular features of each
critical point according to the values of the parameters:
\begin{itemize}
\item $P_0=(0,0)$. The eigenvalues for this critical point are:
\begin{equation}
\mu_H=-\frac{4\omegl+\sigl}{2\omegl+\sigl+1},\;\;\;\;\;
\mu_x=-\frac{2\omegl+2\sigl+3}{2\omegl+\sigl+1}.
\end{equation}
Thus, we have that the critical point is a saddle point for models
whose parameters are between the lines $4\omegl+\sigl=0$ and
$2\omegl+2\sigl+3=0$ whereas it is an attractor node if the
parameters are in the external region. Notice that this critical
point does not exist when $2\omegl+\sigl+1=0$, that corresponds to
Case Ic in which the separatrix $x_s$ is placed at the origin.
When $4\omegl+\sigl=0$ we have that $\mu_H=0$ and the critical
point becomes a critical line because $x=0$ is a singular point
irrespectively of the value of $H$.

\item $P_\pm=(x_c^\pm,0)$. For these critical points the
eigenvalues can be expressed as $\mu_H=-(2x_c^\pm+3)$ and
$\mu_x=-2x_c^\pm$. Then, if the critical point is positive we have
an attractor node whereas it behaves as a repelling node if
$x_c^\pm<-3/2$. Finally, in the range $-3/2<x_c^\pm<0$ we get a
saddle point. These ranges correspond to the regions in the
parameter space showed in Fig. \ref{criticalpoints1}. The
eigenvalues $\mu_x^\pm$ vanish when the corresponding critical
point is located at the origin, i.e., $2\sigl+2\omegl+3=0$. Then,
we can have three different cases depending on the sign of
$2\omegl+\sigl+1$, namely:
\renewcommand{\labelenumi}{\textit{\roman{enumi}})}

\begin{enumerate}
\item $2\omegl+\sigl+1>0$. In this case $\mu_x^+=0$ and
$\mu_x^-=-6(2\omegl+\sigl+1)<0$.

\item $2\omegl+\sigl+1=0$. This is the case IIIb described above
in which both critical points are located at the origin and, as a
consequence, $\mu_x^+=\mu_x^-=0$.

\item $2\omegl+\sigl+1>0$. This case is the opposite to $i$),
i.e., $\mu_x^-=0$ and $\mu_x^+=-6(2\omegl+\sigl+1)>0$.
\end{enumerate}
On the other hand, the eigenvalues $\mu_H^\pm$ vanish in the case
that the critical points are such that $x_c^\pm=-3/2$ which is
satisfied for models with $\sigl+4\omegl=3/4$. In this case we
also obtain three possibilities in terms of the sign of
$12\omegl-15/2$ as follows:
\renewcommand{\labelenumi}{\textit{\roman{enumi}})}
\begin{enumerate}
\item $12\omegl -15/2>0$. In this case we have $\mu_H^-=0$ and
$\mu_H^+=24\omegl-15>0$.

\item $12\omegl -15/2=0$. In this case we have
$\mu_H^-=\mu_H^+=0$.

\item $12\omegl -15/2<0$. In this case we have $\mu_H^+=0$ and
$\mu_H^-=24\omegl-15<0$.
\end{enumerate}

\item Separatrix $x=x_s$. For values of $x$ close to $x_s$, i.e.,
$x=x_s+\delta$ with $\delta\rightarrow 0$ the equations become:
\begin{eqnarray}
\frac{dH}{dN}&\simeq& \mu_s\frac{H}{\delta^2}\\
\frac{dx}{dN}&\simeq& -\mu_s\frac{1}{\delta}
\end{eqnarray}
with
\begin{equation}
\mu_s=9(2\omegl+\sigl+1)
\left[12\omegl^2+\sigl(4+3\sigl)+2\omegl(5+6\sigl)\right].
\end{equation}
Then, the separatrix will attract the trajectories of the phase
map for models in which $\mu_s>0$ whereas for models with
$\mu_s<0$ the trajectories will go away from $x_s$. Note that this
is true for both sides of the separatrix.

\item $x\rightarrow\pm\infty$. In this case we have that
$\frac{dx}{dN}\sim -x^2$ so that the trajectories will always
approach from $+\infty$ and will move away to $-\infty$, i.e., the
region with large values of $x$ $+\infty$ repel the trajectories
of the phase map whereas they become attracted by the region with
$x\rightarrow-\infty$ irrespectively of the values of the
parameters in the action. This means that the region with large
positive values of $x$ is always unstable for any choice of
$\omegl$ and $\sigl$ and, on the contrary, the region with
negative large values of $x$ is always stable.

\end{itemize}

\begin{table}[h!]
\begin{center}
\footnotesize{
\begin{tabular}{|c|c|c|c|c|}
 \hline & & & &\\
 Region & $P_0$  & $P_-$ & Separatrix & $P_+$ \\
& & & &\\
\hline \hline & & & &\\
 I&  Attractor node & Saddle point $\mu_x>0$ & Repelling & Attractor node  \\
& & & &\\
\hline & & & &\\
 II& Saddle point $\mu_x>0$  & Attractor node & Repelling & Attractor node\\
& & & &\\
\hline & & & &\\
 III& Saddle point $\mu_x>0$  & -- & Attractor & --\\
& & & &\\
\hline & & & &\\
 IV& Saddle point $\mu_x<0$  & -- & Repelling & -- \\
& & & &\\
\hline & & & &\\
 V& Saddle point $\mu_x<0$  & Repelling node & Attractor & Repelling node\\
& & & &\\
\hline & & & &\\
 VI& Attractor node  & Repelling node & Attractor & Repelling node\\
& & & &\\
\hline & & & &\\
 VII& Attractor node  & Repelling node & Attractor & Saddle point $\mu_x>0$\\
& & & &\\
\hline & & & &\\
 VIII& Saddle point $\mu_x>0$  & Repelling node & Attracting & Attractor node\\
& & & &\\
\hline & & & &\\
 IX&  Saddle point $\mu_x<0$ & Repelling node & Repelling & Attractor node\\
& & & &\\
\hline & & & &\\
 X& Saddle point $\mu_x<0$  & Saddle point $\mu_x>0$ & Repelling & Attractor node\\
& & & &\\
\hline& & & &\\
 XI& Saddle point $\mu_x>0$  & Saddle point $\mu_x>0$ & Attractor & Attractor node\\
& & & &\\
\hline& & & &\\
 XII& Attractor node  & Saddle point $\mu_x>0$& Attractor & Saddle point $\mu_x>0$\\
& & & &\\
\hline& & & &\\
 XIII& Attractor node  & -- & Repelling & --\\
& & & &\\
\hline
\end{tabular}}
\vspace{0.3cm} \caption{\small{In this table we summarize the
features of the phase map for the different regions shown in Fig.
\ref{criticalpoints1}. When a given critical point is a saddle
point we give the sign of the eigenvalue corresponding to $x$ so
that we can know whether the trajectories approach the critical
point (negative eigenvalue) or move away from it (positive
eigenvalue) along the $x$-direction.}}\label{tablaclas}
\vspace{0.1cm}
\end{center}
\end{table}

\newpage

\subsection{Accelerating solutions}
\begin{figure}[h!]
\begin{center}
{\epsfxsize=14cm \epsfbox{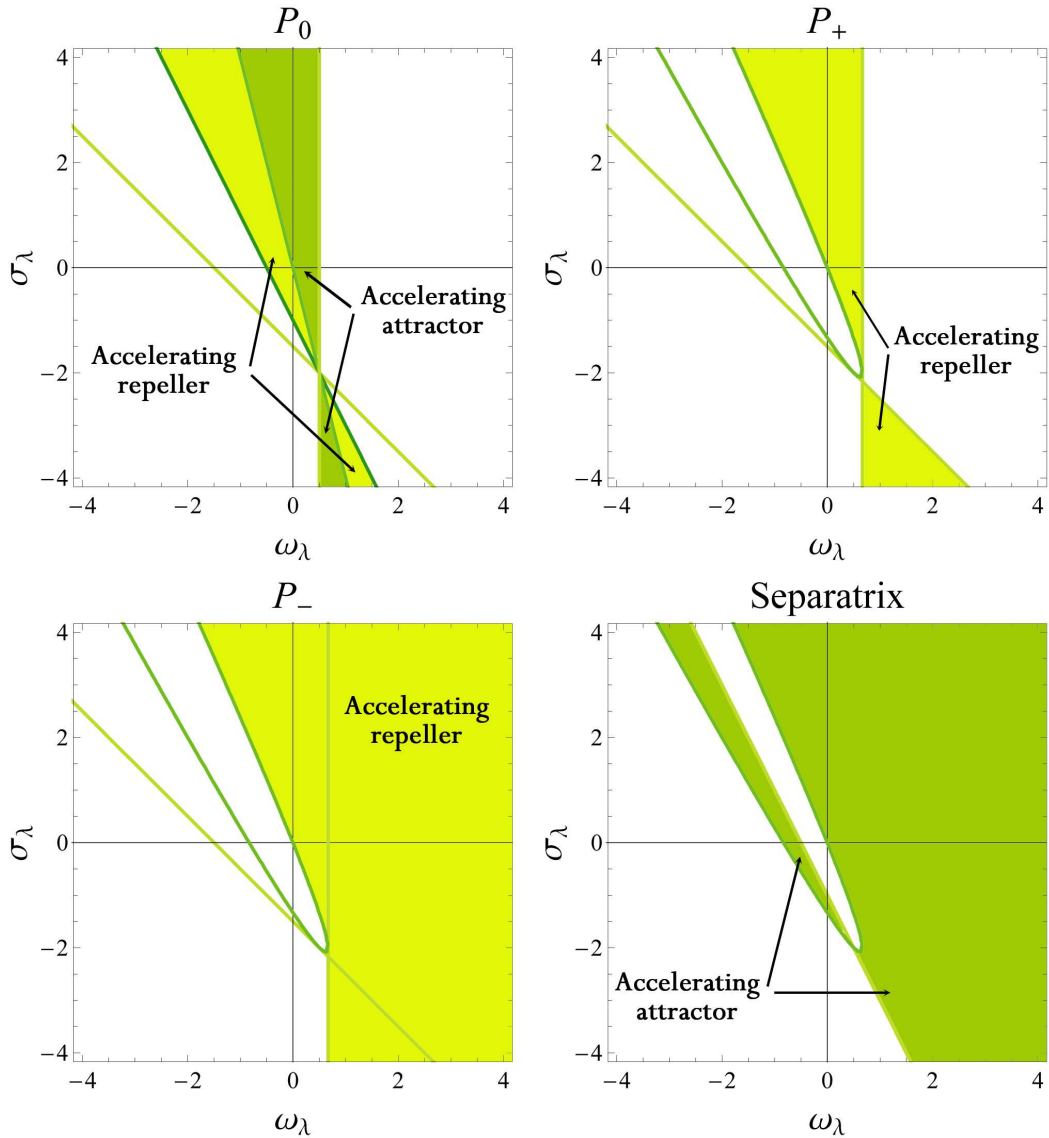}}\caption{\small In this plot
we show the regions in which we get accelerated solutions in a
universe dominated by the vector field for each critical point.
The darkest shaded regions are those in which the accelerated
solutions are attractors. We can see that neither for $P_+$ nor
for $P_-$ we can obtain attracting accelerated solutions.
}\label{LTacc}
\end{center}
\end{figure}

In this section we shall enumerate the necessary conditions for a
vector-tensor model to lead to accelerating solutions. To that end
we shall express the equation of state in terms of the field
variable $x$:
\begin{equation}
w=-\frac{(4\omegl+2\sigl+1)x^2-2(2\omegl-\sigl-3)x-3(2\omegl+\sigl+1)
(2\omegl-\sigl-3)}{\left[x+3(2\omegl+\sigl+1)\right]^2}\label{asympw}
\end{equation}
Notice that this equation of state only depends on $x$ and not on
the Hubble parameter $H$. The models with accelerating solutions
will be those in which $w$ evolves towards $w<-1/3$. Moreover, as
this work is intended to find models in which the vector field
could play the role of dark energy, we shall demand that the
accelerated phase is an attractor. To that end, we shall look at
the equation of state for all the possible attracting places in
the phase map as well as in the repelling ones. Notice that, as
the equation of state does not depend on $H$ but only on $x$ we
only need to require attractor or repelling properties with
respect to $x$. For instance, if a critical point is a saddle
point but with the trajectories going towards $x\rightarrow
x_{crit}$ ($\mu_x<0$) it will be considered as an attractor and
the opposite for a repelling point. Now, we shall study the
existence of accelerating regimes in the phase map:

\begin{itemize}
\item $P_0$. As we pointed out above, this critical point will be
an attractor in models with $\mu_x<0$. On the other hand, the
equation of state for this critical point is
\begin{equation}
w_{P_0}=\frac{2\omegl-\sigl-3}{3(2\omegl+\sigl+1)}
\end{equation}
From this expression we see that $P_0$ corresponds to an
accelerated phase ($w_{P_0}<-1/3$) if the following condition
holds:
\begin{equation}
\frac{2\omegl-1}{2\omegl+\sigl+1}<0\label{acccondP0}
\end{equation}
which is satisfied for models in which either
$1<2\omegl<-(1+\sigl)$ or $-(1+\sigl)<2\omegl<1$. The
corresponding region is shown in Fig. \ref{LTacc} and, in that
figure, we see that there exists a region in which $P_0$ is an
attractor and gives rise to accelerated expansion simultaneously.

\item $P_\pm$. For these critical points, the equation of state
becomes:
\begin{equation}
w_{P_\pm}=-(8\omegl+4\sigl+3)\pm
\frac{4}{3}\sqrt{6(5\omegl+2\sigl)+9(2\omegl+\sigl)^2}
\end{equation}
Notice that if the condition of accelerated expansion is satisfied
for $P_+$, then, it is also satisfied for $P_-$ since $w_{P_+}\geq
w_{P_-}$. The models that lead to accelerated expansion are shown
in Fig. \ref{LTacc}. However, in the same figure, we see that
neither of the critical points $P_\pm$ behaves as an attractor in
the region where we get accelerated expansion.

\item Separatrix $x=x_s$. In this case the equation of state will
evolve either to $+\infty$ or to $-\infty$ as the trajectory
approaches the separatrix. The interesting case here is when
$w\rightarrow-\infty$ so that we get acceleration. For
$x=x_s+\delta$ with $\delta\rightarrow 0$, the equation of state
becomes:
\begin{equation}
w_s\simeq-\frac{2\mu_s}{3\delta^2}
\end{equation}
In Fig. \ref{LTacc} we show the region in which the equation of
state goes to $-\infty$, corresponding to the condition $\mu_s>0$.
Notice that this condition also guarantees that the separatrix
behaves as an attractor so all the cases in which the sepratrix is
an attractor give rise to accelerated expansion.

\item $x\rightarrow \pm\infty$. As we showed above, $x=-\infty$ is
always an attractor whereas $x=+\infty$ is always a repelling
point. Moreover, as $x$ approaches $\pm\infty$ the equation of
state is given by:
\begin{equation}
w(x\rightarrow\pm\infty)=-(4\omegl+2\sigl+1)
\end{equation}
so that it gives rise to accelerated expansion in models with
$2\omegl+\sigl>-1/3$. This case is interesting for the vector
field to drive an inflationary era because it could start with a
large value of $x$ and, as $x=\infty$ repels the trajectories, it
would be forced to evolve towards smaller values of $x$ until it
reaches either $P_+$ or $P_0$. Moreover, such an evolution can
lead to accelerated expansion as long as the condition
$2\omegl+\sigl>-1/3$ holds.

\end{itemize}

After having obtained the general conditions necessary to have accelerated expansion, we shall study the particular solutions in which the scale factor evolves
as a power of the cosmic time, i.e., $a\propto t^p$ and we shall  get some analytical solutions. For this expansion law, the vector field evolves
according to (\ref{solRM}), although, in this case, the parameter $p$ must be
determined from Einstein equations. When we introduce
(\ref{solRM}) in (\ref{asympFeq}) we obtain that the vector field
must take a constant value given by:
\begin{equation}
A_0^\infty=\frac{\pm 1}{\sqrt{8\pi G\lambda(3+2\omegl+2\sigl)}}
\end{equation}
Notice that this value only makes sense for
$\lambda(3+2\omegl+2\sigl)>0$, which can always be fulfilled by a
suitable choice of the parameter $\lambda$. Moreover, we also
obtain that $p$ is given by:
\begin{equation}
p=\frac{1+2\omegl+\sigl}{4\omegl+\sigl}
\end{equation}
With this expression we can calculate the deceleration parameter:
\begin{equation}
q_\infty\equiv-\frac{\ddot{a}a}{\dot{a}^2}=\frac{2\omegl-1}{2\omegl+\sigl+1}
\end{equation}
This deceleration parameter must be negative in order to have an
accelerated expansion, but this is indeed the same condition that
we found above in (\ref{acccondP0}) when studying the critical
point $P_0$. This was expected because the Hubble expansion rate
goes to zero for a power law expansion and this imposes that $A_0$
has to be constant which means that $x=0$. Thus, we have just
obtained nothing but the analytical solutions for the trajectories
approaching $P_0$ along the $H$-axis. Notice that this is, indeed,
the only critical point attracting trajectories with accelerated
expansion and having a finite value for the vector field equation of
state.

\section{Transition from matter domination to vector domination:
late-time accelerated solutions}

In previous sections we have studied a universe completely
dominated by the vector field and obtained the necessary
conditions to have accelerated solutions. In this section,
however, we shall consider the case in which the universe contains
matter in addition to the vector field and study the circumstances
under which we can get a transition from a matter dominated
universe to an accelerated phase provided by the vector field so
that the vector field can play the role of dark energy. To do so,
we shall proceed as in previous sections, i.e., we shall obtain
the corresponding autonomous system and identify attracting
solutions in which the vector field eventually dominates the
energy content of the universe and has equation of state smaller
that $-1/3$.

Before going on with the study of the autonomous system, we remind
that a necessary condition to have a candidate to dark energy is
that the energy density associated to the vector field decays
slower than that of a pressureless fluid in the matter dominated
epoch. This requirement guarantees the dominance of the vector
field at late times so that it can drive the expansion of the
universe. According to Table 1, such models are those whose
parameters satisfy the condition
\begin{equation}
2\omegl-\sigl<3.\label{ltA0dom}
\end{equation}
We shall take this condition as a necessary requirement for the
model to be able to play the role of dark energy.

The system of equations must be modified by introducing the matter
contribution to the Friedman equation so that:
\begin{equation}
3H^2=8\pi G(\rho_{A_0}+\rho_M).
\end{equation}
Moreover, we have a new equation provided by the energy
conservation of the matter fluid:
\begin{equation}
\dot{\rho}_M+3H\rho_M=0
\end{equation}
As we did in the previous sections, we shall introduce the field
variable $x\equiv\frac{d\ln A_0}{d\ln a}$ and the matter energy
density will be described by the density parameter
$\Omega_M\equiv\frac{\rho_M}{3H^2}$. In terms of these variables,
we can obtain the following autonomous system:
\begin{eqnarray}
\frac{dH}{dN}&=&3\frac{(2\omegl+\sigl)x^2-2(4\omegl+\sigl)x-3(4\omegl+\sigl)
(2\omegl+\sigl+1)+F_H\Omega_M}
{\left[x+3(2\omegl+\sigl+1)\right]^2
-\left[9(2\omegl+\sigl)^2+6(5\omegl+2\sigl)\right]\Omega_M}H\\
\frac{dx}{dN}&=&-\frac{\left[x^2+6(2\omegl+\sigl+1)x+3(2\omegl+2\sigl+3)\right]
\left[x(x+3(2\omegl+\sigl+1))+F_x\Omega_M\right]}
{\left[x+3(2\omegl+\sigl+1)\right]^2
-\left[9(2\omegl+\sigl)^2+6(5\omegl+2\sigl)\right]\Omega_M}\nonumber\\
\frac{d\Omega_M}{dN}&=&3\frac{\left[(4\omegl+2\sigl+1)x^2-(4\omegl-2\sigl-6)x
-3(2\omegl+\sigl+1)(2\omegl-\sigl-3)\right]}{\left[x+3(2\omegl+\sigl+1)\right]^2
-\left[9(2\omegl+\sigl)^2+6(5\omegl+2\sigl)\right]\Omega_M}(1-\Omega_M)\Omega_M\nonumber
\label{autsysM}
\end{eqnarray}
where we have defined:
\begin{eqnarray}
F_H&=&-\frac{1}{2}\left[(4\omegl+2\sigl+1)x^2-2(2\omegl-\sigl-3)x
+6(4\omegl+\sigl)(2\omegl+\sigl)+9(2\omegl-1)\right]\nonumber\\
F_x&=&-\frac{3}{2}\left[(4\omegl+2\sigl+1)x-2\omegl+\sigl+3\right].
\end{eqnarray}
These equations are supplemented by the following constraint
provided by the Friedman equation:
\begin{equation}
\frac{1}{3}\lambda
A_0^2\left[x^2+6(2\omegl+\sigl+1)x+3(2\omegl+2\sigl+3)\right]+\Omega_M=1
\end{equation}
As before, this relation will determine the sign of the parameter
$\lambda$ in order to fulfill the condition $\lambda
\left[x^2+6(2\omegl+\sigl+1)x+3(2\omegl+2\sigl+3)\right]>0$.

The equation of state for the vector field is given in this case
by:
\begin{equation}
w=-\frac{(4\omegl+2\sigl+1)x^2-2(2\omegl-\sigl-3)x-3(2\omegl+\sigl+1)(2\omegl-\sigl-3)}
{\left[x+3(2\omegl+\sigl+1)\right]^2-\left[9(2\omegl+\sigl)^2+6(5\omegl+2\sigl)\right]\Omega_M}\label{asympw}
\end{equation}

\begin{figure}[hb!]
\begin{center}
{\epsfxsize=15cm \epsfbox{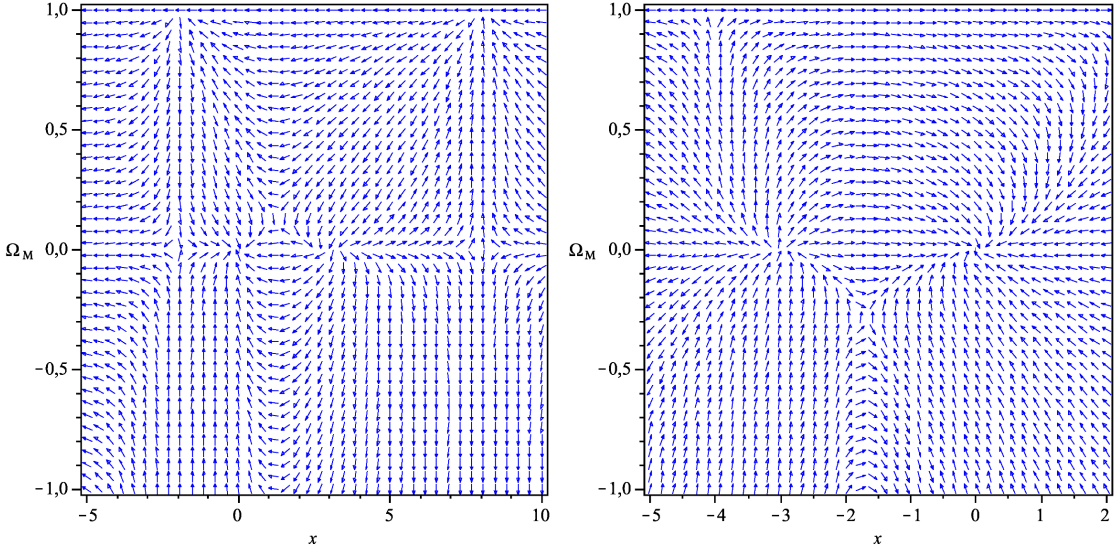}} \caption{\small{In these
two plots we show two examples of phase maps corresponding to the
cases when the two critical points are present (left) and when
they are not (right), or, equivalently, when the separatrix is
open from above or from below.}}\label{mapaomega}
\end{center}
\end{figure}

 The sections $\{x,\Omega_M\}$ of the phase map do not depend on
$H$. Notice that, as expected, the equations for $\frac{dH}{dN}$
and $\frac{dx}{dN}$ reduce to (\ref{autsys}) when $\Omega_M=0$ so
that all the critical points analyzed in the previous sections are
also critical points of (\ref{autsysM}) with $\Omega_M=0$, which
is the interesting case here because this means that the vector
fields eventually dominates the energy of the universe. Apart from
these critical points with $\Omega_M=0$ we also have critical
points with $\Omega_M=1$ which correspond to situations where
matter drives the universe expansion. As we are interested in
obtaining solutions leading to late-time accelerated expansion
driven by the vector field, we shall study just the critical
points with $\Omega_M=0$ so that the critical values of $x$ and
$H$ are the same as those studied in the vector dominance case.
However, the features of the critical points may change because of
the presence of matter. Therefore, for each critical point, we
shall identify the necessary conditions for the corresponding
critical point to be an attractor with respect to $\Omega_M$ and
$x$ and, then, compute the equation of state in the critical
point. This is possible because the equations for $\Omega_M$ and
$x$ does not depend on $H$. Notice that the condition for
$\Omega_M=0$ to be an attractor will, in general, differ from the
condition given in (\ref{ltA0dom}) obtained by imposing that the
energy density of the vector field grows with respect to that of
matter. This is so because to achieve that condition we assumed
that the amount of matter was initially dominant with respect to
that of the vector field and, however, for $\Omega_M=0$ to be an
attractor, such a condition does not need to be satisfied.
Finally, it is interesting to remark that the region with
$\Omega_M>0$ is disconnected from that with $\Omega_M<0$ because
the trajectories are always tangent to the plane $\Omega_M=0$
which ensures that the energy density of matter remains positive
as long as it is initially positive. See Fig. \ref{mapaomega} to
have an idea of how the phase maps look like.

Let us analyze then each critical point for this case:

\begin{itemize}
\item $P_0=(0,0)$. For this critical point, the linearized system
becomes:
\begin{eqnarray}
\frac{dH}{dN}&\simeq&
-\frac{4\omegl+\sigl}{2\omegl+\sigl+1}H\nonumber\\
\frac{dx}{dN}&\simeq& -\frac{2\omegl+2\sigl+3}{2\omegl+\sigl+1}
\left[x+\frac{2\omegl-\sigl-3}{2(2\omegl+\sigl+1)}\Omega_M\right]\nonumber\\
\frac{d\Omega_M}{dN}&\simeq&
\frac{2\omegl-\sigl-3}{2\omegl+\sigl+1}\Omega_M\nonumber
\end{eqnarray}
The eigenvalues for this system are the same as those of the
vector dominance case plus
$\mu_{\Omega_M}=(2\omegl-\sigl-3)/(2\omegl+\sigl+1)$, which
determines the stability of the solutions with
$\Omega_M\rightarrow0$. Therefore, the analysis proceeds exactly
the same as before with the supplementary condition
$\mu_{\Omega_M}<0$ ensuring the late-time domination of the vector
field. Notice that, as commented above, this condition is not the
same as that given in (\ref{ltA0dom}). Indeed, this supplementary
condition happens not to reduce the region of the parameter space
in which we get attracting solutions with accelerated expansion,
i.e., all the models indicated in Fig. (\ref{LTacc}) corresponding
to these type of solutions have indeed $\Omega_M=0$ as an
attractor.

\item $P_\pm=(x_\pm,0)$. In this case the linearized system
becomes diagonal with the same eigenvalues for $x$ and $H$ as in
the vector dominance case, i.e., $\mu_H=-(2x_c^\pm+3)$ and
$\mu_x=-2x_c^\pm$. Moreover, the eigenvalue for $\Omega_M$ is
given by $\mu_{\Omega_M}=4x_c^\pm+3$. Hence, as in the case of
$P_0$ the stability analysis is the same as that already performed
above, although we must impose the condition $4x_c^\pm+3<0$ so
that the vector field eventually dominates. However, when we
impose the latter condition we find that these critical points
happen not to be attractors for any value of the parameters.
\end{itemize}

So far, we have seen that the presence of a matter fluid only
affects the solutions in the sense that the parameter space is
restricted to that region in which $\Omega_M=0$ is an attractor,
otherwise the vector field would never dominate and we cannot
produce late-time acceleration. In other words, the features of
the critical points remain the same as those studied in the vector
domination case, although only the cases in the allowed region are
admissible. The novelties appear when studying the separatrix:
\begin{itemize}
\item Separatrix. In this case the separatrix is no longer given
by $x=x_s$, but by the parabola:
\begin{equation}
\Omega_M=\frac{\left[x+3(2\omegl+\sigl+1)\right]^2}
{\left[9(2\omegl+\sigl)^2+6(5\omegl+2\sigl)\right]}
\end{equation}
Notice that the vertex of this parabola always lies on the
$x$-axis, i.e., in the vertex we always get $\Omega_M=0$ so that
it will be interesting to have solutions attracted by it. Whether
this parabola is open from above or below in the $(x,\Omega_M)$
plane depends on the sign of
$9(2\omegl+\sigl)^2+6(5\omegl+2\sigl)$ which also determines the
existence of $x_c^\pm$ so that if the critical points $x_c^\pm$
exist the parabola goes up and if they do not exist the parabola
goes down. Since negative values of $\Omega_M$ are physically
unreasonable, an open from below parabola does not represent a
proper separatrix for the physically admissible region of the
phase map.

Close to the vertex of the separatrix, i.e., for:
\begin{eqnarray}
\Omega_M&=&\frac{\left[x+3(2\omegl+\sigl+1)\right]^2}
{\left[9(2\omegl+\sigl)^2+6(5\omegl+2\sigl)\right]}+\delta_{\Omega_M}
\\\nonumber\\
x&=&-3(2\omegl+\sigl+1)+\delta_x\label{closevertex}
\end{eqnarray}
with $\delta_{\Omega_M},\delta_x\ll 1$ the autonomous system
becomes:
\begin{eqnarray}
\frac{dH}{dN}&\simeq&3\frac{2\omegl+\sigl+1}{\delta_{\Omega_M}}H
\nonumber\\
\frac{dx}{dN}&\simeq&3(2\omegl+\sigl+1)\frac{\delta_x}{\delta_{\Omega_M}}
\nonumber\\
\frac{d\Omega_M}{dN}&\simeq&
2\frac{2\omegl+\sigl+1}{3(2\omegl+\sigl)^2+2(5\omegl+2\sigl)}
\frac{\delta_x^2}{\delta_{\Omega_M}}\nonumber
\end{eqnarray}
In the previous expressions, $\delta_{\Omega_M}$ parametrizes the
separation to the separatrix and is positive (negative) for points
above (below) it, whereas $\delta_x$ gives the separation on the
right ($\delta_x>0$) or on the left ($\delta_x<0$) from the vertex
of the parabola. From the equation for $dH/dx$ we see that the
trajectories will not be able to cross the vertex because $dH/dN$
becomes singular at that point. On the other hand, it is easy to
see from the equations for $dx/dN$ and $d\Omega_M/dN$ that the
vertex will always act as an attractor for the trajectories
approaching from one side of the parabola so that, in the cases in
which the parabola is open from above, there always exist
trajectories that are attracted by the vertex of the separatrix.
Whether the trajectories approaching the vertex are those going
from above or below the parabola is determined by the sign of
$2\omegl+\sigl+1$ as follows:
\renewcommand{\labelenumi}{\textit{\roman{enumi}})}
\begin{enumerate}
\item $2\omegl+\sigl+1>0$. In this case the trajectories
approaching from below are attracted towards the vertex whereas
those solution contained in the region above the separatrix are
repelled by the vertex.

\item $2\omegl+\sigl+1<0$. In this case, the solutions in the
region above the separatrix are attracted towards the vertex
whereas the trajectories below the parabola go away from the
vertex.
\end{enumerate}

To study the cases when the parabola is open from below we shall
analyze the autonomous systems for
$x+3(2\omegl+\sigl+1)=\delta_x\ll 1$ and $\Omega_M\ll 1$. In that
case we obtain
\begin{eqnarray}
\frac{dx}{dN}&\simeq& 3(2\omegl+\sigl+1)\frac{\delta_x}{\Omega_M} \nonumber\\
\frac{d\Omega_M}{dN}&\simeq& 6(2\omegl+\sigl+1)
\end{eqnarray}
Thus, only if $(2\omegl+\sigl+1)<0$ the vertex of the separatrix
is an attractor when the separatrix is contained in the region
with $\Omega_M<0$.

\begin{figure}[hb!]
\begin{center}
{\epsfxsize=12cm \epsfbox{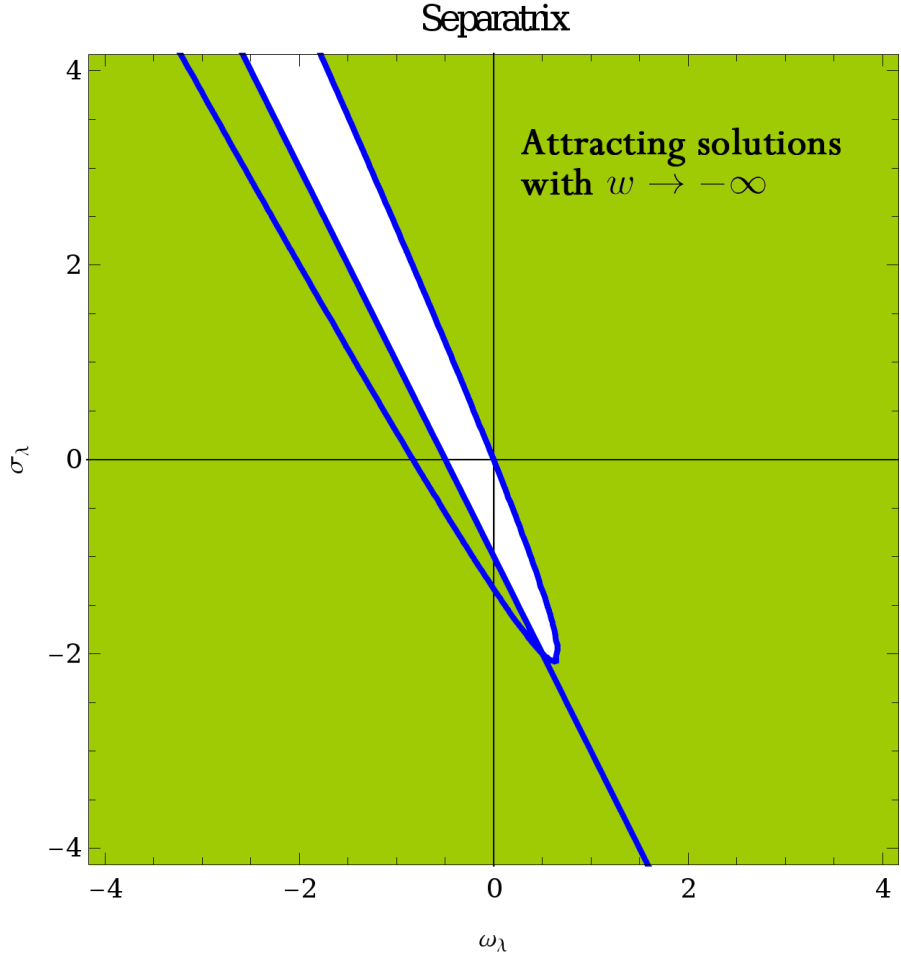}} \caption{\small{ In this
plot we show the regions where we have the vertex of the
separatrix attracting trajectories with accelerated expansion
(green shaded region) and where there are not solutions attracted
by the vertex in which the expansion is accelerated (white
region). We see that, in most of the parameter space, we have that
the vertex attracts some trajectories in which the equation of
state of the vector field diverges evolving towards $-\infty$. The
curve plotted in the graph separates those models in which the
separatrix is open from above (outer region) and those in which
the separatrix is open from below (inner region). Notice that
these regions coincide with those in which the critical points
$P_\pm$ exist (separatrix open from above) and they do not
(separatrix open from below), as explained in the main text.}}
\end{center}
\end{figure}

Finally, it remains to study the behavior of the equation of state
as the trajectory approaches the vertex of the separatrix. If we
use the parametrization given in (\ref{closevertex}) again, we
obtain that the equation of state becomes:
\begin{equation}
w\simeq 2\frac{2\omegl+\sigl+1}{\delta_{\Omega_M}}
\end{equation}
Therefore, if the trajectory approaches the vertex from above the
separatrix, the equation of state will evolve towards $+\infty$
($-\infty$) as long as $2\omegl+\sigl+1$ is positive (negative).
On the contrary, when the trajectory goes to the vertex from below
the separatrix the equation of state of the vector field goes to
$+\infty$ ($-\infty$) as long as $2\omegl+\sigl+1$ is negative
(positive).

Then, if the parabola is open from above
($3(2\omegl+\sigl)^2+2(5\omegl+2\sigl)>0$), irrespectively of the
sign of $2\omegl+\sigl+1$ we have that the vertex acts as an
attractor for some trajectories with solutions whose equation of
state goes to $-\infty$. On the other hand, when the parabola is
open from below ($3(2\omegl+\sigl)^2+2(5\omegl+2\sigl)<0$), only
when $2\omegl+\sigl+1$ is negative the vertex acts as an attractor
and, in that case, the equation of state in the vertex goes to
$-\infty$.

\end{itemize}

Finally, we would like to comment on the existence of certain
models in which we can have critical points with $\Omega_M\neq
0,1$. Those critical points can been found from (\ref{autsysM}) by
solving $\frac{d\Omega_M}{dN}=0$ with respect to $x$ and for
arbitrary values of $\Omega_M$ and, then, obtain the corresponding
critical value for $\Omega_M$ from the equation $\frac{dx}{dN}=0$.
In these critical points, one generally gets $H=0$. The explicit
expressions for these critical points are:
\begin{eqnarray}
x_{\Omega_M}^\pm&=&\frac{2\omegl-\sigl-3\pm
\sqrt{2(2\omegl-\sigl-3)\left[2(5\omegl+2\sigl)+3(2\omegl+\sigl)^2\right]}}
{4\omegl+2\sigl+1}\\
\Omega_M^\pm
&=&-\frac{x_{\Omega_M}^\pm\left[x_{\Omega_M}^\pm+3(2\omegl+\sigl+1)\right]}{F_x(x_{\Omega_M}^\pm)}
\end{eqnarray}
Therefore, only models in which
$(2\omegl-\sigl-3)\left[2(5\omegl+2\sigl)+3(2\omegl+\sigl)^2\right]$
is positive can contain these type of critical points. Moreover,
this condition does not guarantee the existence of physically
admissible values for $\Omega_M$ because one could, in principle,
obtain both positive and negative values for $\Omega_M$. However,
these critical points cannot lead to accelerated solutions because
the equation of state for the vector field in such points is
identically zero, i.e., it behaves as a dust fluid.

To summarize the results of this section, we have shown that the
features of the critical points for the case when we have matter
in addition to the vector field remain unaffected, but the
behavior of the separatrix presents novelties and, generally, in
all the models we shall have attracting solutions with future
singularities. Finally, we have shown that solutions in which
$\Omega_M$ goes to some values different from $0$ and $1$ are such
that the equation of state of the vector field goes to zero, i.e.,
it asymptotically behaves as a matter fluid.

\section{Local gravity tests}
The viability of any alternative theory of gravity is subject to
its agreement with Solar System experiments, which provide very
tight constraints on the so-called Parametrized-Post Newtonian
(PPN) parameters. These parameters are a set of quantities that
characterize most of gravity theories at small scales and are
extremely useful to measure deviations from GR. For the action
(\ref{Action2}), the PPN parameters are given by \cite{Will}:
\begin{eqnarray}
\gamma&=&\frac{1+4\omega
A^2\left(1+\frac{2\omega+\sig}{\epsilon}\right)}
{1-4\omega A^2\left(1-\frac{4\omega}{\epsilon}\right)}\nonumber\\
\beta&=&\frac{1}{4}(3+\gamma)+\frac{1}{2}\Theta
\left[1+\frac{\gamma(\gamma-2)}{G}\right]\nonumber\\
\alpha_1&=&4(1-\gamma)\left[1+2\epsilon\Delta\right]+16\omega A^2\Delta a\nonumber\\
\alpha_2&=&3(1-\gamma)\left[1+\frac{4}{3}\epsilon\Delta\right]+8\omega
A^2\Delta a-2\frac{bA^2}{G}\nonumber\\
\alpha_3&=&\xi=\zeta_1=\zeta_2=\zeta_3=\zeta_4=0\label{PPNVT}
\end{eqnarray}
with:
\begin{eqnarray}
&\Theta&=\frac{(1-4\omega A^2)(2\epsilon+\sig-2\omega)}{(1-4\omega
A^2)2\epsilon+32\omega^2A^2}\nonumber\\
&\Delta&=\frac{1}{2A^2\sig^2-2\epsilon
\left[1-4A^2(\omega+\sig)\right]}\nonumber\\
&a&=2\epsilon(1-3\gamma)+2\sig(1-2\gamma)\nonumber\\
&b&=\left\{\begin{array}{c}
(2\omega+\sig)\left[(2\gamma-1)(\gamma+1)+\Theta(\gamma-2)\right]\\
-(2\gamma-1)^2(2\omega+\sig+\lambda)\left(1-\frac{2\omega+\sig+\lambda}{\lambda}\right)\;\;\;\;\;\;\;\;\;\;\;\lambda\neq0\\\\
\;\;\;\;\;\;\;\;\;\;\;\;\;\;0\;\;\;\;\;\;\;\;\;\;\;\;\;\;\;\;\;\;\;\;\;\;\;\;\;\;\;\;\;\;\;\;\;\;\;\;\;\;\;\;\;\;\;\;\;\;\;\;\;\;\;\;\lambda=0
\end{array} \right.\nonumber
\end{eqnarray}
Moreover, the effective Newton's constant is defined as:
\begin{eqnarray}
G_{eff}&\equiv& G\left[\frac{1}{2}(\gamma+1)+6\omega
A^2(\gamma-1)-2A^2\sig(1+\Theta)\right]^{-1}.
\end{eqnarray}
In the above expressions we have assumed $G_{eff}=1$ and $A$ is
the value of the vector field at Solar System scales (in units of
$4\pi G$). The parameters $(\gamma,\beta)$ are usually called the
static PPN parameters and measure the space-curvature produced by
a unit mass and the degree of nonlinearity relative to GR
respectively. The parameter $\xi$ measures effects of preferred
location whereas $\alpha_i$ have to do with preferred frame
effects. Finally, $\alpha_3$ and $\zeta_i$ are non-vanishing for
theories in which the conservation of total momentum is violated.
In GR, the PPN parameters are such that
$\gamma-1=\beta-1=\alpha_1=\alpha_1=\alpha_3=\xi=\zeta_1
=\zeta_2=\zeta_3=\zeta_4=0$. On the other hand, for a general
vector-tensor theory we see that there are neither preferred
location effects nor violation of the total momentum
conservations. However, these theories typically lead to preferred
frame effects (as expected because of the presence of a vector
field) as well as deviations from GR for the static PPN
parameters. Current observational limits on the PPN parameters
impose very stringent limits on modified gravity theories because
they do not allow much deviation from GR, i.e., GR agrees with
local gravity tests with very good precision \cite{Will}:
\begin{eqnarray}
\gamma-1&\lsim&2.3 \times10^{-5}\nonumber\\
\beta-1&\lsim& 2.3\times10^{-4}\nonumber\\
\alpha_1&\lsim& 10^{-4}\nonumber\\
\alpha_2&\lsim& 10^{-4}\; (10^{-7})\label{PPNlimits}
\end{eqnarray}
In order to obtain constraints on the vector field from these
limits we linearize the PPN parameters given in (\ref{PPNVT}) as
follows:
\begin{eqnarray}
\gamma-1&\simeq&\frac{4\omega}{\epsilon}\left[2(\epsilon-\omega)+\sig\right]A^2\nonumber\\
\beta-1&\simeq&\frac{(2\omega-\sig)(2\epsilon-2\omega+\sig)(4\epsilon-2\omega+\sigl)}
{4\epsilon^2}A^2\nonumber\\
\alpha_1&\simeq&\frac{16\omega(2\epsilon+\sig)}{\epsilon}A^2\nonumber\\
\alpha_2&\simeq&\left\{\begin{array}{c}
\left[(2\omega+\sig)^2\left(\frac{1}{\epsilon}-\frac{2}{\lambda}\right)-4\sig\right]A^2\;\;\;\;\;\;\;\;\;\;\;\lambda\neq0\\\\
\frac{4\omega}{\epsilon}\left[2(\epsilon+\omega)+\sig\right]A^2\;\;\;\;\;\;\;\;\;\;\;\;\;\;\;\;\;\;\;\;\;\;\;\lambda=0
\end{array} \right.
\end{eqnarray}
Therefore, we can set that, typically, the vector field at the
Solar System scale will be constrained to be $A\lsim 10^{-2}$, for
models in which all the parameters are order unity. Let's remark
that this value of the vector field does not need to coincide with
its cosmological value.

On the other hand, the linearized Newton's constant is given by:
\begin{equation}
G\simeq
1-\left[\frac{(2\omega-\sig)^2}{\epsilon}+4(\sig-\omega)\right]A^2
\end{equation}
Then, if we use the existent limits on its time-variation
$\dot{G}/G\lsim 10^{-13}$yr$^{-1}$ together with the constraints
on the vector field obtained above, we can also set bounds on the
cosmological time variation of the vector.

Although a general vector-tensor theory will be constrained by the
aforementioned limits, there is a number of models whose
parameters satisfy certain relations so that some of the PPN
parameters could be identical to those of GR and, thus, pass the
limits given in (\ref{PPNlimits}). Indeed, in \cite{viable} is
shown that there exist a total of 6 models which are
indistinguishable from GR by means of local gravity tests for any
value of the background vector field, namely: $\{\sig=0,\epsilon,
\lambda\}$, $\{\sig=-4\lambda=-4\epsilon\}$,
$\{\sig=-3\lambda=-2\epsilon\}$ and $\{\sig=m\epsilon,$ with
$m=0,-2,-4\}$. Moreover, in that work, a detailed analysis of the
stability of these models is performed.

\section{Conclusions and discussion}
In this work we have developed a general study of the cosmological
evolution of a vector field non-minimally coupled to gravity and
without potential terms. We have given the evolution of this
vector field in terms of the parameters of the theory for the
different epochs of the expansion history of the universe, namely:
inflation, radiation dominated era and matter dominated era. We
have shown that it is possible to obtain a wide variety of
behaviors for the evolution of the vector field by suitable
choices of the parameters. In particular, we have obtained
conditions for the parameters so that the vector field energy
density grows or decays with respect to that of the dominant
component. Moreover, conditions to have scaling evolution have
also been calculated.

The case of a universe dominated by the temporal component of the
vector field has been studied in detail. We have obtained an
autonomous system describing the evolution of the Hubble expansion
rate and the vector field. The general features of the phase map
have been given and all the critical points have been
appropriately characterized. For those points that act as
attractors we have obtained the general conditions under which the
vector field gives rise to accelerated expansion.

To study the viability of these theories as dark energy candidates
we have performed an analysis of the field equations together with
Einstein's equations when the universe contains matter in addition
to the vector field. Then, we have identified solutions which
allow a transition from matter-domination to vector-domination
with accelerated expansion. We have also shown that these models
generally contain future singularities in which the equation of
state diverges and that most of the models can give rise to
periods of accelerated expansion.

In addition to the general results commented above, we have also
found a wide variety of particular model examples with interesting
properties, thus: models with late-time attractors with
$\Omega_M\neq 0, 1$ all with equations of state for the vector
field resembling that of non-relativistic matter. This type of
models could provide vector dark matter candidates. We have also
found models with early time accelerated solutions which are
unstable thus giving rise to possible finite inflationary periods.
On the other hand, the dark energy model proposed in
\cite{cosmicvector} corresponds to the parameters
$\epsilon=-1/4,\lambda=-1/2,\;\sigma=1,\;\omega=0$. Concerning the
temporal component, we have that $\sigl=-2$ and $\omegl=0$ and, if
we look at Fig. \ref{A0clasification}, we see that it lies in the
line of scaling behavior during radiation (what allows to solve
the coincidence problem in a natural manner) and in the growing
region for the matter era, in agreement with the result found in
\cite{cosmicvector}. Moreover, it belongs to the region I of the
classification summarized in Table \ref{tablaclas} for the vector
dominance case so that $P_0$ and $P_+$ behave as attractors
whereas $P_-$ and the separatrix repel the trajectories. The
singularity found in \cite{cosmicvector} corresponds to the
trajectories approaching the separatrix in the case when matter is
present. This is also in agreement with the results of this paper
because we have that the separatrix for this particular model is
given by $\Omega_M=\frac{1}{12}(x-3)^2$ so that it is open from
above. Moreover, we have that $2\omegl+\sigl+1=-1<0$ so that,
according to the discussion about the separatrix in section 5, the
trajectories above the separatrix will be attracted by its vertex.
Then, as the vector field evolves as $\propto t^\alpha$ with
$\alpha=\frac{1}{6}(-3+\sqrt{3})$ in the matter dominated era, we
have that the initial conditions are given by $\Omega_M\simeq 1$
and $x\simeq\frac{3}{2}\,\alpha\simeq0.69$ so that the
cosmological solution corresponds to one of the trajectories
approaching the vertex of the separatrix and, therefore, leading
to a future singularity. Finally, the electromagnetic dark energy
model proposed in \cite{EMDE} has parameters $\omega=\sigma=0,\;
\lambda=1/6,\; \epsilon=-1/4$ and coincides with the only model
whose temporal component has constant energy density for all the
cosmological epochs (see Fig. \ref{A0clasification}). Moreover, if
we look at Fig. \ref{VDregion1} we see that it belongs to Case
IIIa in which both critical points $P_+$ and $P_-$ coalesce and
they coincide with the sepratrix. In \cite{EMDE} it is shown that
this model, indeed, behaves as a cosmological constant
irrespectively of the Hubble expansion rate so that it leads to a
de-Sitter universe once it dominates over the matter component.

In the last section of the paper we have given the expressions for
the PPN parameters of the general vector-tensor theory and used
current limits on the PPN parameters to constrain the value of the
vector field at small (Solar System) scales, which has been found
to be $A_\odot\lsim 10^{-2}$. However, one cannot, in principle,
use this bound to put limits on the cosmological value of the
vector field because both values do not need to coincide, unless
the power spectrum of the vector field happens to be
scale-invariant. In the general case, one would need to know the
mechanism that originated the primordial power spectrum of the
vector field so that we can know the expectation value of the
vector field at Solar System scales and, thus, compare to the
experimental limits.

We would like to remark that, throughout this work, we have
focused on the viability of vector-tensor theories as candidates
for dark energy just for the first stage, i.e., we have given the
general conditions under which the homogeneous part of a
vector-tensor theory can lead to late-time acceleration. However,
to propose a serious candidate one should also study the
perturbations of the corresponding model and check the presence of
instabilities both at classical and quantum levels. The absence of
classical instabilities is required in order not to have
exponentially growing modes that could spoil the predictions of
the model for the zero-mode. The issue of quantum instabilities
has to do with the existence of modes with negative energy
(ghosts) so that non-linear interactions of the field might
produce an unlimited number of such particles. The existence of
such instabilities has already been studied in \cite{viable} for
those vector-tensor theories which are indistinguishable from GR
at small scales by considering perturbations in both the vector
field and the metric. In \cite{Peloso}, it was shown the existence
of instabilities for the longitudinal component of the vector
field in some particular cases. In \cite{Armendariz}, a detailed
treatment on the stability of the vector field perturbations for
general vector-tensor theories with a potential term was
performed. However, a complete study by considering metric
perturbations in addition to vector perturbations remains to be
done, although such a study is far from straightforward because of
the large number de cases involved. In any case, as suggested in
\cite{forms} where general N-forms are studied, the existence of
singular classical solutions is related to the presence of ghosts
so that one would expect that most of the vector-tensor models
present instabilities (at least at the quantum level) because, as
shown in this work, the existence of singular solutions at the
classical level is a common feature in these models.

As a final comment, although throughout this work we have focused
on vector fields as candidates for dark energy, we would like to
mention that, because of their generality, the results given in
the paper can also be used in other cosmological contexts in which
vector fields could play a relevant role.

{ \large{\em Acknowledgments:}} This work has been  supported by
Ministerio de Ciencia e Innovaci\'on (Spain) project numbers FIS
2008-01323 and FPA 2008-00592, UCM-Santander PR34/07-15875,
UCM-BSCH GR58/08 910309 and MEC grant BES-2006-12059.

\end{document}